\documentclass[10pt,journal,compsoc]{IEEEtran}
\ifCLASSOPTIONcompsoc
  \usepackage[nocompress]{cite}
\else
  \usepackage{cite}
\fi
\ifCLASSINFOpdf
\else
\fi
\usepackage[utf8]{inputenc}
\usepackage[T1]{fontenc}

\ifCLASSOPTIONcompsoc
  \usepackage[nocompress]{cite}
\else
  \usepackage{cite}
\fi

\usepackage{amsmath}
\usepackage{amssymb}
\usepackage{amsthm}
\newtheorem{theorem}{Theorem}
\newtheorem{lemma}{Lemma}
\newtheorem{corollary}{Corollary}

\newcommand{\cmark}{\checkmark}
\newcommand{\xmark}{\ensuremath{\times}}

\usepackage{graphicx}
\usepackage{stfloats}          % bottom double-column floats
\ifCLASSOPTIONcompsoc
  \usepackage[caption=false,font=normalsize,labelfont=sf,textfont=sf]{subfig}
\else
  \usepackage[caption=false,font=footnotesize]{subfig}
\fi

\usepackage{tikz}
\usetikzlibrary{
    shapes.geometric,
    arrows.meta,
    positioning,
    calc,
    fit,
    decorations.pathreplacing,
    shadows.blur,
    backgrounds
}

\usepackage{algorithm}
\usepackage{algpseudocode}

\usepackage{booktabs}
\usepackage{array}

\usepackage{enumitem}
\usepackage{balance}           % \balance before \bibliography to even last-page columns

\usepackage[colorlinks=true, citecolor=blue, linkcolor=blue, urlcolor=blue]{hyperref}

\begin{document}
\title{Bridging the Epistemic Gap in the Invisible Internet: Extended Mathematical Modeling and Empirical Characterization of I2P Topology}
\author{Siddique~Abubakr~Muntaka,~\IEEEmembership{Graduate~Student~Member,~IEEE,}
        Jacques~Bou~Abdo,~\IEEEmembership{Senior~Member,~IEEE,}
        and~Liaquat~Hossain%
\IEEEcompsocitemizethanks{%
\IEEEcompsocthanksitem S. A. Muntaka and J. Bou Abdo are with the School of
Information Technology, University of Cincinnati, Cincinnati, OH 45221 USA.\protect\\
E-mail: muntaksr@mail.uc.edu; bouabdjs@ucmail.uc.edu
\IEEEcompsocthanksitem L. Hossain is with the College of Science, Long Island
University, Brooklyn, NY 11201 USA.\protect\\
E-mail: MdLiaquat.Hossain@liu.edu}% <-this % stops an unwanted space
}

\markboth{IEEE Transactions on Network Science and Engineering}%
{Muntaka \MakeLowercase{\textit{et al.}}: Bridging the Epistemic Gap in the Invisible Internet}

\IEEEtitleabstractindextext{%
\begin{abstract}
The Invisible Internet Project (I2P) is a decentralized, peer-to-peer network underpinning secure and adversarial communications with major implications for cybersecurity and national security. Its global topology emerges from thousands of independent peer-selection decisions, producing a hub-dominated, complex structure that resists external observation. Prior models based on protocol documentation failed to capture real-world behavior. We close this gap with an Extended Mathematical Model (EMM) derived from production code and validated on a six-router, geographically distributed testbed over thirty days. Results reveal limited router visibility ($\sim$25\% of nodes), relativistic tier allocations, strong Fast-tier tunnel preference, and overrepresented High Capacity-tier exploratory tunnels. The model accurately predicts emergent topological asymmetries and defines the adversarial resources required to compromise privacy, providing a principled foundation for cyber deterrence and secure decentralized network design.
\end{abstract}

}

\maketitle

\IEEEdisplaynontitleabstractindextext

% \fi
\IEEEpeerreviewmaketitle

% ===== IEEE under-review notice — ARXIV VERSION ONLY =====
\noindent\begingroup
\setlength{\fboxsep}{7pt}
\fcolorbox{red}{red!5}{%
\begin{minipage}{0.95\columnwidth}\footnotesize
\textbf{\textcolor{red}{Notice:}} This work has been submitted to the IEEE for
possible publication. Copyright may be transferred without notice, after which
this version may no longer be accessible.
\end{minipage}}
\endgroup\par\medskip

\section{Introduction}\label{sec:introduction}


\IEEEPARstart{T}{he}
Invisible Internet Project (I2P) is a peer-to-peer overlay network with major implications for national security, cybersecurity, cyber warfare, and homeland security, as well as for net neutrality and freedom of speech~\cite{muntaka2025resilience}. For example, its anonymous infrastructure has been exploited as a command-and-control backbone for global botnet operations~\cite{georgoulias2023market}, while simultaneously serving as a censorship-resistant channel for users in restrictive regimes~\cite{de2019invisible}. The empirical study of I2P confronts a categorical obstacle: I2P is, by architectural design, volatile, unpredictable, and resistant to measurement \cite{muntaka2025mapping}. Continuous node churn \cite{muntaka2025mapping}, locally executed peer-selection decisions, and a selectively replicated network database (NetDB) ensure that no external vantage point commands a structurally complete or temporally stable full view of the system. Researchers applying empirical methods have produced datasets that are incomplete, temporally inconsistent, and systematically biased toward the observable periphery of a network whose most consequential properties are precisely those least visible from outside \cite{muntaka2025mapping} \cite{muntaka2025resilience}. This limitation is not resolvable by denser sampling; it is a structural consequence of the system's design. The appropriate scientific response is therefore not a refinement of empirical methodology but a change in paradigm: from observation to derivation. Network science, grounded in the tradition from Euler's 1736 graph-theoretic proof \cite{euler1741solutio} through Erdős and Rényi's probabilistic graph theory \cite{erdos1959publicationes}, Watts and Strogatz's small-world formalization \cite{watts1998collective}, and Barabási and Albert's scale-free derivation from preferential attachment \cite{barabasi1999emergence}, provides precisely this alternative: macroscopic structural properties are not observed but derived, as necessary analytical consequences of local interaction rules. Applied to I2P, this paradigm shifts attention from crawling what the network appears to be toward formally characterizing what it must be, given the peer-selection algorithms, floodfill election thresholds, and tunnel construction.

To the best of our knowledge, within the full body of I2P topology literature, a single prior contribution has pursued this paradigm. Bou Abdo and Hossain \cite{abdo2023modeling} produced the only published work that analytically models I2P as an emerging network topology, grounded in network science and complex systems theory rather than external observation, establishing for the first time that I2P's structural properties are derivable rather than merely cataloguable. The work of Bou Abdo and Hossain~\cite{abdo2023modeling} is therefore the direct intellectual predecessor of the present contribution. Its foundational limitation, however, is that its assumptions were derived from I2P's public documentation rather than from the source code that deployed routers actually execute. I2P's documented and implemented behaviors diverge in ways that are architecturally significant: peer-selection criteria, bandwidth tier thresholds, floodfill eligibility conditions, and tunnel construction parameters in production deployments differ materially from their documented counterparts. A model grounded in documentation-based assumptions is therefore a model of a network that does not exist in production, producing formally valid but empirically misgrounded conclusions. This gap, between specification and implementation, is the precise methodological problem the present work resolves.
This paper advances the analytical modeling of I2P through three sequential contributions, each grounded in implementation analysis. The first demonstrates formally, drawing on implementation 
parameters extracted through systematic reverse engineering of 
I2P's production source code in prior work~\cite{muntaka2026systemic}, 
that the assumptions of Bou Abdo and Hossain are inconsistent 
with deployed router behavior. The second proposes corrective, implementation-validated assumptions 
and establishes their empirical validity through controlled testbed 
experimentation using instrumented I2P routers in floodfill and 
standard configurations, grounded in the implementation parameters 
extracted in~\cite{muntaka2026systemic}. The third and principal contribution derives from these corrected assumptions and, using the formal machinery of probability theory, tier modeling, and graph-theoretic degree distribution analysis, an Extended Mathematical Model (EMM) of I2P's emerging topology that predicts router degree distributions and tier-stratified selection probabilities as analytical consequences of its implementation-grounded mechanisms. 

The significance of these contributions is defined by the problem they resolve across more than two decades of I2P research; to the best of our knowledge, no work has provided a formal, mathematically grounded characterization of the network's anonymity guarantees. Anonymity in I2P is probabilistic, a function of compromised node fraction, floodfill concentration, path diversity, and degree distribution bottlenecks. These are quantities derived only from a topology model, not from empirical snapshots. The EMM is the first work to satisfy this precondition, allowing empirical approximations to be replaced by formal mathematical theorems regarding anonymity. The necessity of this contribution is supported by the fact that I2P has already established a substantial footprint in the global cybersecurity threat landscape. I2P has been used as a backbone to dark web command-and-control (C2) systems, where operators of botnets can maintain a stable channel of communication that resists traffic analysis and takedown at the infrastructure level \cite{georgoulias2023market}. The I2Ninja botnet is the best-documented I2P-based C2 example; however, malware families such as Dyre, Zeus, Citadel, Ramnit, and SpyEye have also used the anonymous overlay to provide structural redundancy against domain seizure and law enforcement disruption \cite{Maor2014, jeong2016longitudinal, yasui2024spot}. Outside of criminal activity, I2P may be deployed to anonymize the leaking of intellectual property, stolen national security information, and the organization of attacks on critical infrastructure \cite{cyberpress_ratatouille_i2p_2026}. Nation-state actors engaged in offensive cyber operations may use these anonymous overlays as strategic resources. They recognize that weakening the linkage between network-layer activity and physical attribution allows them to engage in sustained, deniable operations, achieving a level of structural unaccountability to which no human-layer procedure can compare. It is against this background that a formal topology model has explicit strategic implications. Cyber deterrence \cite{rid2015attributing} requires communicating the cost of an attack credibly and formally: the EMM makes tractable the adversarial resource thresholds required to compromise I2P's privacy guarantees, expressed as functions of controlled router fraction, floodfill compromise rate, and path diversity collapse conditions. Cyber attribution, legally and technically contested in anonymity-network operations, becomes analytically grounded when the structural configurations under which privacy guarantees degrade are formally characterized. Both capabilities are enabled by the EMM and unattainable without it.

%The new How I2P Works subsection in INTRO

\begin{figure*}[htb]
\centering
\includegraphics[width=0.90\textwidth]{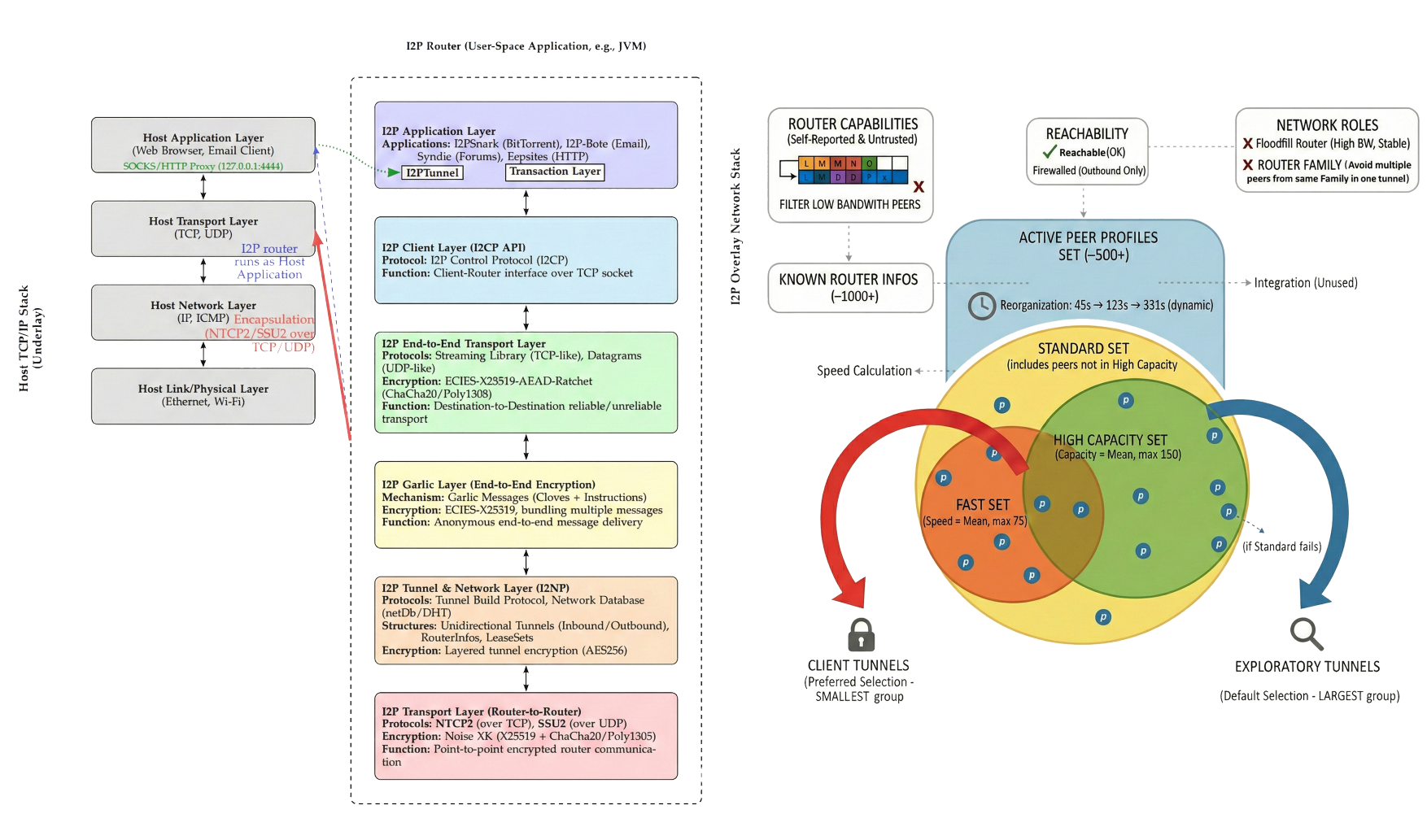}
\caption{%
  \textbf{Left:} The layered architecture of the I2P (Invisible Internet Project) overlay network,
  showing the protocol stack from applications to the transport layer, and its relationship to the
  underlying host TCP/IP stack. The I2P router implements garlic routing with unidirectional tunnels,
  using modern cryptography for end-to-end and transport encryption.
  \textbf{Right:} I2P Dynamic Peer Profiling and Selection. Routers derive an Active Peer Profiles Set from NetDB RouterInfos, re-profiling peers periodically into Fast, High Capacity, and Standard tiers based on observed performance medians (not self-reported capabilities).}
\label{fig:i2p_arch_profile}
\end{figure*}

\subsection{What is I2P and how it works}
\label{sec:i2p_overview}

I2P is a user-space, multi-layered protocol stack, as depicted in the left panel of Fig.~\ref{fig:i2p_arch_profile}, that creates an overlay network running over the host's TCP/IP infrastructure. The stack is organized into six layers. At the lowest layer, the \textit{I2P Transport Layer} provides point-to-point encrypted router-to-router communication using two
protocols: NTCP2 (over TCP) and SSU2 (over UDP), NTCP2 uses the Noise XK handshake with X25519 key agreement 
and ChaCha20/Poly1305 authenticated encryption; SSU2 employs 
a similar Noise-based session establishment adapted for UDP 
with token-based retry mechanisms. Immediately
above sits the \textit{I2P Tunnel and Network Layer}, which implements the I2P Network Protocol
(I2NP). This layer manages the construction of unidirectional tunnels that carry all I2P traffic
and maintains the distributed Network Database (NetDB) via a Kademlia-based distributed hash table
(DHT), storing \textit{RouterInfo} descriptors and \textit{LeaseSets} that map destinations to their
inbound tunnel endpoints. Tunnel payloads are protected by layered AES-256 encryption, where each
intermediate hop decrypts exactly one layer and forwards the remainder, acquiring no knowledge of
the origin, destination, or overall tunnel length beyond its two immediate neighbors.

The \textit{I2P Garlic Layer} extends classical onion routing~\cite{dingledine2004tor} by bundling
multiple encrypted messages, termed \textit{cloves}, into a single \textit{garlic message}, using
ECIES-X25519 for asymmetric key agreement. Delivering several cloves in one transmission obscures
the correspondence between individual messages and their senders, providing enhanced resistance to
traffic analysis compared to the single-message approach of onion routing~\cite{chao2024systematic}.
The layer above, the \textit{I2P End-to-End Transport Layer}, exposes two delivery abstractions to
applications: a TCP-like Streaming Library for reliable ordered delivery and a UDP-like Datagram
service for unreliable delivery, both encrypted end-to-end with ECIES-X25519-AEAD-Ratchet
(ChaCha20/Poly1305). The \textit{I2P Client Layer} exposes these services to client applications
through the I2P Control Protocol (I2CP) API over a local TCP socket, while the \textit{I2P
Application Layer} hosts I2P-native services, including I2PSnark (BitTorrent), I2P-Bote (email),
Syndie (forums), and Eepsites (HTTP hidden services accessible as \texttt{*.i2p} hostnames).

Unlike Tor's bidirectional circuits, I2P creates separate inbound and outbound tunnels for each
communication session~\cite{timpanaro2012bird}, which provides greater resistance to certain
correlation attacks, though at the expense of increased resource utilization. Tunnels are rebuilt on average every ten minutes~\cite{schimmer2009peer}, and construction decisions are executed locally based on queries made by each router to its own partial copy of the NetDB. There are two functionally distinct tunnel types that control the network traffic. The application traffic is transported over \textit{client tunnels}, which are built with strict performance requirements. The \textit{Exploratory tunnels} also facilitate network discovery and NetDB maintenance and are built with a focus on path diversity. The peer-selection policy used for each type of tunnel varies greatly, and this functional differentiation has direct implications for the emergent topology of the network, implications that are formalized for the first time within this paper.

Central to both peer construction processes is the ongoing profiling system illustrated in the right panel of Fig.~\ref{fig:i2p_arch_profile}. Every router has an Active Peer Profiles Set, which is based on a Known RouterInfo pool in its local NetDB, usually containing hundreds of profiled peers~\cite{muntaka2026systemic}. Active routers are categorized into performance tiers by the system depending on locally measured data of throughput and latency. The capabilities of latency and throughput are not self-reported~\cite{herrmann2011privacy}. The upper echelon of the routers (peers or nodes within the garlic network) belong to the \textit{High Capacity set} (capacity $\geq$ mean, capped at 150 peers). Within the High Capacity set is another group called the \textit{Fast set} (speed \(\geq\) mean of the High Capacity pool, capped at a hard ceiling of 75 peers), which contains the absolute highest-performing nodes across both throughput and speed. Any other non-failing, accessible peers fall under the thresholds called \textit{Standard set}, which makes up the largest set of the group in any active profiles of a router on the network. The nested structure \(\gamma_s \subseteq g_{1s} \subseteq
\omega_s\) is therefore determined entirely by local measurement. Since each router computes these
thresholds independently over its own partial network view, the tier composition observed by
router \(s_1\) need not agree with that of router \(s_2\) for the same underlying peer. Profile
reorganization operates on a schedule scaled with router uptime: 45 seconds at startup, 123
seconds during warm-up, and 351 seconds during stable operation~\cite{i2p_peermanager_source}.

\begin{figure}[H]
\centering
\includegraphics[width=0.40\textwidth]{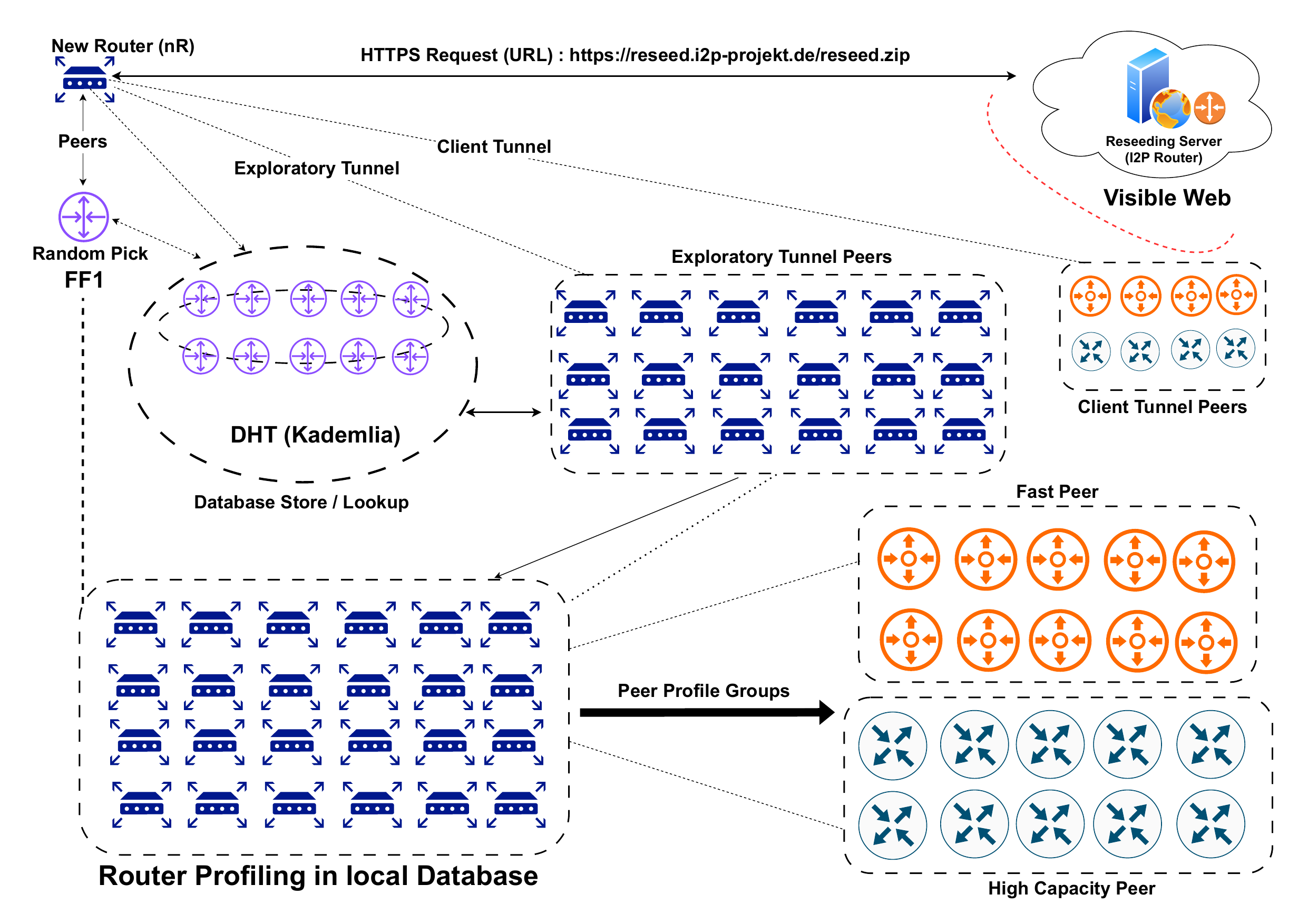}
\caption{The process of a new I2P router joining the network and forming tunnels.}
\label{fig:i2pinitiation}
\end{figure}

A new I2P router joins the network through the deterministic bootstrapping sequence illustrated in
Fig.~\ref{fig:i2pinitiation}. The process begins with \textit{reseeding}, where the new router
downloads RouterInfo entries from public HTTPS servers. It then performs NetDB lookups via
Kademlia-based DHT queries to floodfill routers (high-bandwidth, stable peers responsible for
storing and forwarding NetDB entries) to progressively expand its local peer database.

Once a sufficient number of RouterInfos have been accumulated and profiled, the router constructs
exploratory tunnels to further expand its network view and participates in ongoing NetDB
propagation. Client tunnels are subsequently established once the profiling cycle has populated
the Fast tier with sufficient peers to satisfy construction requirements, completing the
transition from network isolation to active participation. Each of these locally executed peer-selection decisions, when aggregated across thousands of
routers, collectively shapes the macro-level topology whose structural properties determine I2P's
anonymity guarantees, resilience under adversarial targeting, and the C\&C infrastructure it
shelters. No individual router observes this global structure; each operates within the partial network slice its local NetDB affords. Thus, deriving the network structure analytically is more rigorous than approximating it through empirical observation, which is inherently incomplete~\cite{muntaka2026fifty}.

The remainder of the paper is structured in the following way. Section~\ref{litreview} presents the technical background and architecture of I2P, measurement studies, and peer profiling mechanisms, and discusses the interplay between micro-level peer selection choices and macro-level topology formation, with a critique of the literature on anonymity network modeling by comparing top-down and bottom-up approaches while enumerating the weaknesses of documentation assumptions in I2P. In Section~\ref{methodology}, we outline the three stages of our methodology, which includes examining the assumptions in the single novel analytical model of the authors in~\cite{abdo2023modeling}. We then examine I2P using their mathematical model assumptions to describe the network alongside the empirical test using the framework provided in Fig.~\ref{fig:cloudswarm}. Section~\ref{EMM} presents the construction of the Extended Mathematical Model incorporating tier-stratified selection, empirical validation of its corollaries, and the predicted emergent global topology. Section~\ref{sec:discussion} discusses the structural implications of the EMM predictions, including comparison with scale-free network models, security consequences of the degree asymmetry. Section~\ref{conclusion} presents
conclusions and future research directions.

\section{Related Work} \label{litreview}

The analytical study of I2P's network topology confronts a structural constraint that runs through the entire body of existing work: the most consequential properties of the network resist direct observation. An anonymity guarantee is not an observable metric. Adversarial leverage thresholds are not observables. The degree distribution governing which routers bear the structural weight of the entire network's tunnel construction is an emergent phenomenon characteristic of complex systems, born of thousands of locally executed peer-selection decisions.

The literature surveyed in this section has approached this problem almost entirely through an observational paradigm. Measurement studies characterize what the network appears to be. However, the cybersecurity demands on I2P research require formally characterizing what the network must be, given the precise peer-selection algorithms and performance-tier thresholds its routers actually execute. System science and network theory provide precisely this alternative. Rooted in the tradition from Euler's graph-theoretic proof~\cite{euler1741solutio} through Barab\'{a}si and Albert's scale-free 
derivation~\cite{barabasi1999emergence}, this discipline establishes that macroscopic structural properties in network systems are derived as necessary analytical consequences of local interaction rules. The literature surveyed here has not yet applied this bottom-up approach to I2P with implementation-grounded rigor. This section determines why such a gap is there, what previous literature is capable of within its confines, and what it is incapable of offering without the formal methods or approach as presented in this work.

\subsection{Foundational Concepts in Anonymous Communication}

The first mix networks introduced by Chaum~\cite{chaum1981untraceable} 
established the foundational principle of communication mixing and defined the basic principle of linkage breaking between messages and senders through cryptographic transformation of these messages. Onion routing~\cite{dingledine2004tor} built upon this observation by creating encrypted paths that were layered. Tor uses onion routing through a centralized directory authority that offers a conventional, openly auditable view of all network nodes ~\cite{murdoch2005low}. The fact that centralization is present is what causes the topology of Tor to be observed, rather than being fully emergent, as in I2P. Since all routers in Tor make decisions based on the same directory state, the topology of Tor is treated as an external input to analysis and can be modeled mathematically ~\cite{dingledine2004tor}. 

These characteristics are categorically unavailable with the I2P architecture. I2P is a complex adaptive system that is fully decentralized \cite{muntaka2025optimizing}. Introduced through a distributed network database implemented on the principles of Kademlia \cite{maymounkov2002kademlia}, I2P ensures that any router has only partial, localized network information. The article by De Boer and Breider ~\cite{de2019invisible} provides a comprehensive review of the I2P architecture, revealing design trade-offs of decentralization and censorship resistance over low latency. These trade-offs have precise analytical consequences in systems science. The I2P topology results from collective peer selection by independent routers. To derive such a topology analytically, one would need a bottom-up approach in formalizing those decisions. The decisions should be extracted from the system to formalize them. When I2P has already been studied using top-down techniques, the effect is a partial perspective of the network since it disregards the localized, emergent character of the routing topology in the network. Garlic routing is founded on onion routing, except that a number of messages are encrypted in packages, which are a lot more resilient to traffic analysis~\cite{chao2024systematic}. I2P establishes inbound and outbound tunnels on each session, unlike Tor, which uses bidirectional circuits for a session~\cite{timpanaro2012bird}.

I2P Tunnels regenerate themselves repeatedly with a different set of functional demands~\cite{schimmer2009peer}. Client tunnels handle application traffic and are prioritized by the Fast performance tier. Exploratory tunnels are used to find network peers and are based on a larger population. This functional differentiation generates asymmetric selection on the various router tiers, and is the major generative process that causes heterogeneity in degree distributions. 

However, despite the original design intentions to resist traffic analysis, recent empirical studies demonstrate that I2P hidden services remain highly susceptible to large-scale de-anonymization via live behavior alignment~\cite{wang2025time}. The topology formally derived in I2P provides the structural explanation for this exact vulnerability: an adversary does not need to monitor the entire network, but merely position themselves within this hyper-centralized Fast tier to achieve the visibility required for such attacks. No prior study has applied implementation-grounded formal methods to quantify the network-wide structural consequences of this mechanism.

\subsection{Top-Down Empirical Studies: Measurement and Characterization}

The majority of I2P research has employed top-down empirical methodologies, measuring aggregate network behavior through active crawling and passive observation. While these studies provide ecological validity, their structural limitation is inherent to the paradigm: they answer descriptive questions about observed network state but cannot derive anonymity guarantees, predict topological consequences of algorithmic 
changes, or characterize the emergent structural properties that determine adversarial leverage.

Early empirical data about the I2P growth patterns and node distributions was generated by Liu et al.~\cite{liu2019i2p}. Although they gave a useful overview of the patterns of use, their information was not able to correlate those patterns to the micro-level peer-selection protocols that fuel the formation of the network. Timpanaro et al.~\cite{timpanaro2011monitoring} investigated the floodfill routers of I2P, determining that these high-capacity nodes take up key structural positions. Knowledge about the behavior of floodfills is crucial, but the work of Timpanaro did not use a formal mathematical model to show how the emergent degree distribution was determined by the floodfill election thresholds.

The empirical investigation of the nature of I2P tunnels by Hoang et al.~\cite{hoang2018empirical} revealed that a significant portion of tunnels have fewer than the default recommended number of hops in the protocol. This shows that users choose paths based on performance rather than anonymity specifications. This observation directly raises the risk of using design specifications to conduct security assessments. The gap between what is stated in the protocol recommendations and what actually happens during the construction of the tunnel can cause an overestimation of the network's anonymity guarantees for researchers who use only documentation to study the working process.

Empirical research on I2P peer-selection schemes confirmed 
that selection does not follow a uniform random model. Herrmann and Grothoff~\cite{herrmann2011privacy} demonstrated that there is a systematic bias toward high-performance nodes, which leads floodfill routers to be chosen for tunnels at rates that are significantly higher than their proportion. These findings confirmed empirically that the operational topology of I2P is highly non-random. The correct analytic reaction is hence a paradigm shift toward analytical modeling, with the application of network theory to trace these localized biases to their macroscopic structural endpoints.

\subsection{Security Analysis, Cyber Warfare, and the Threat Landscape}

I2P security analysis faces a structural issue that can no longer be addressed using the empirical tradition. The aspects of an anonymity guarantee in which the guarantee becomes weak, the network thresholds above which adversarial compromise is structurally important, and the network structures that protect malicious infrastructure are all topological functions. Without a formal topology model, security analyses can prove the feasibility of attacks but cannot bound their overall effects.

This structural gap has direct consequences for contemporary 
cyber warfare operations. I2P provides the basic infrastructure used by dark web command-and-control (C2) systems, which allow botnet operators to have stable communication channels that are resistant to infrastructure-level takedown~\cite{georgoulias2023market}. In the case of advanced persistent threats and nation-state actors, anonymous overlays are used to obtain structural redundancy and deniability in attack activities~\cite{yasui2024spot, cyberpress_ratatouille_i2p_2026}. Cyber deterrence~\cite{rid2015attributing} demands credible and formal communication of the cost of an attack. The use of empirical snapshots of I2P leads defenders to flawed security assumptions, since it does not contain the mathematical tipping points where privacy guarantees fail in the face of adversarial resource investment. These deterrence thresholds have to be formalized by a bottom-up method of topology modeling.

Yin and He~\cite{yin2019i2p} explored the differentiation between I2P and ordinary Internet traffic with the help of machine learning. Their model presupposed prior knowledge of all the traffic and did not model the behavior of paths in a dynamic manner. This is because the structural positions of the routers that have the highest load of tunnel construction can be found only using a degree distribution model obtained under system science. Sybil attacks on peer-to-peer anonymity networks have been studied by Borisov et al.~\cite{borisov2007denial}, and practical Sybil attacks against I2P have been shown by Egger et al.~\cite{egger2013practical} by controlling measured performance metrics. An adversary that manages to put controlled nodes into the Fast tier of I2P acquires excessive structural power. Yet, without a formal topology model, these prior studies can only demonstrate that such structural manipulation is empirically feasible; they lack the mathematical machinery to quantify the resulting degree asymmetry or calculate the exact bounds of the adversary's advantage. That precise quantification of adversarial leverage cannot be derived without the formal analytical model this work provides.

\subsection{Bottom-Up Analytical Modelling: From Documentation to Implementation}

The appropriate analytical response to I2P's resistance to empirical characterization is the application of the paradigm network theory established for complex systems. This paper derives the network's degree distribution as a necessary analytical consequence of its implementation-grounded peer-selection rules. Classical models fail to capture I2P's topology because their generative mechanisms are incorrect. I2P's selection is tier-stratified, bounded by hard population caps enforced in source code, and conditioned on locally observed performance measurements. 

Bou Abdo and Hossain~\cite{abdo2023modeling} produced the first analytical model linking I2P's micro-level peer-selection mechanisms to macro-level network structure. Their model formalized peer-selection probabilities to predict expected network degree distributions. This established for the first time that I2P's structural properties are derivable rather than merely cataloguable. This pioneering work is the direct intellectual predecessor of the present contribution. Its limitation resides solely in its epistemic grounding. The assumptions driving the model were derived from I2P's public documentation rather than the deployed source code. I2P's documented and implemented behaviors diverge in ways that are architecturally significant. A model grounded in documentation-based assumptions produces formally valid but empirically misgrounded conclusions. It is exactly this purpose that the present work serves by bridging the gap between specification and implementation using rigorous formal methods.

\subsection{Synthesis: Determining the Research Gap}

\begin{table*}[!t]
\centering
\caption{Comparative Positioning of the Proposed EMM Against Representative I2P Topology and Security Studies}
\label{tab:related_work}
\footnotesize
\setlength{\tabcolsep}{5pt}
\renewcommand{\arraystretch}{1.35}
\begin{tabular}{@{}>{\raggedright\arraybackslash}p{3.2cm} >{\raggedright\arraybackslash}p{2.3cm} >{\raggedright\arraybackslash}p{2.5cm} c c c c@{}}
\toprule
\textbf{Study} & \textbf{Paradigm} & \textbf{Parameter Basis} &
\shortstack{\textbf{Formal}\\\textbf{Topology}\\\textbf{Model}} &
\shortstack{\textbf{Tier-Stratified}\\\textbf{Bounded}\\\textbf{Selection}} &
\shortstack{\textbf{Implementation}\\\textbf{Validated}} &
\shortstack{\textbf{Adversarial}\\\textbf{Leverage}\\\textbf{Quantified}} \\
\midrule
Liu et al.~\cite{liu2019i2p} & Top-down empirical & Measurement & \xmark & \xmark & \xmark & \xmark \\
Timpanaro et al.~\cite{timpanaro2011monitoring} & Top-down empirical & Measurement & \xmark & Partial\textsuperscript{a} & \xmark & \xmark \\
Hoang et al.~\cite{hoang2018empirical} & Top-down empirical & Measurement & \xmark & \xmark & \xmark & \xmark \\
Herrmann~\& Grothoff~\cite{herrmann2011privacy} & Top-down empirical & Measurement & \xmark & Partial\textsuperscript{b} & \xmark & Partial \\
Egger et al.~\cite{egger2013practical} & Top-down / attack & Measurement & \xmark & \xmark & \xmark & Partial\textsuperscript{c} \\
Bou~Abdo~\& Hossain~\cite{abdo2023modeling} & Bottom-up analytical & Documentation & \cmark & \cmark\textsuperscript{d} & \xmark & \xmark \\
\textbf{This work (EMM)} & Bottom-up analytical & Implementation (source code) & \cmark & \cmark & \cmark & \cmark \\
\bottomrule
\end{tabular}
\\[2pt]
{\footnotesize \raggedright
\textsuperscript{a}Characterizes floodfill routers only, without a degree-distribution model.\quad
\textsuperscript{b}Reports high-performance selection bias but does not model tier caps.\quad
\textsuperscript{c}Demonstrates Sybil feasibility without quantifying degree asymmetry.\quad
\textsuperscript{d}Tiers modeled from documentation, not implementation-enforced caps.\par}
\end{table*}

Table~\ref{tab:related_work} consolidates this comparison, positioning each representative study against the dimensions that define the present contribution. This review demonstrates that there is a methodological gap in I2P studies. Top-down empirical research, such as the works by Liu et al. and Hoang et al.~\cite{liu2019i2p, hoang2018empirical}, employs observational data to answer the question of what the network does. They are unable to predict the way topology would evolve with algorithmic changes or define mathematical anonymity guarantees. The mathematical bottom-up model of Bou Abdo and Hossain~\cite{abdo2023modeling} was the first strategy to use micro-level probabilities to obtain topology at the macro-level. Its weakness is that it is based on assumptions using documentation. Where the implementation is not identical to the specification, predictions made using documentation are not equal to the operational behavior of the network, which may lead to systematic overestimation of the anonymity guarantees and resilience properties of the network.

The missing connection is a model of analysis that is based on validated implementation parameters rather than protocol documentation. The systematic reverse engineering of production I2P source code in~\cite{muntaka2026systemic} retrieved the exact algorithmic logic, threshold values, and hard population limits that define the construction of tunnels in deployed routers. The present paper builds on that basis, with those tested parameters being included in an Extended Mathematical Model whose predictions are verified by controlled testbed experiments.

% ==================
% METHODOLOGY — Section III
% ===================

\section{Methodology} \label{methodology}

The structural reality of the Invisible Internet Project determines the methodological imperative of this paper. Any attempt to characterize the emergent topology of such a system from idealized design documentation introduces systematic analytical errors regarding the structural leverage available to adversaries. 

Documentation defines how a network ought to work, but the manner in which the network is actually formed is defined by operating codebase logic and local routing facts.

Bou Abdo and Hossain~\cite{abdo2023modeling} developed the first mathematical modeling of I2P peer selection. There is, however, a structural drawback to their model in that the underlying assumptions were based upon the interpretation of protocol specifications as opposed to measurements of the implementation deployed. To evolve the analytical modeling of I2P, the necessary methodological advance is to systematically recalibrate these assumptions against operationally verified implementation parameters.

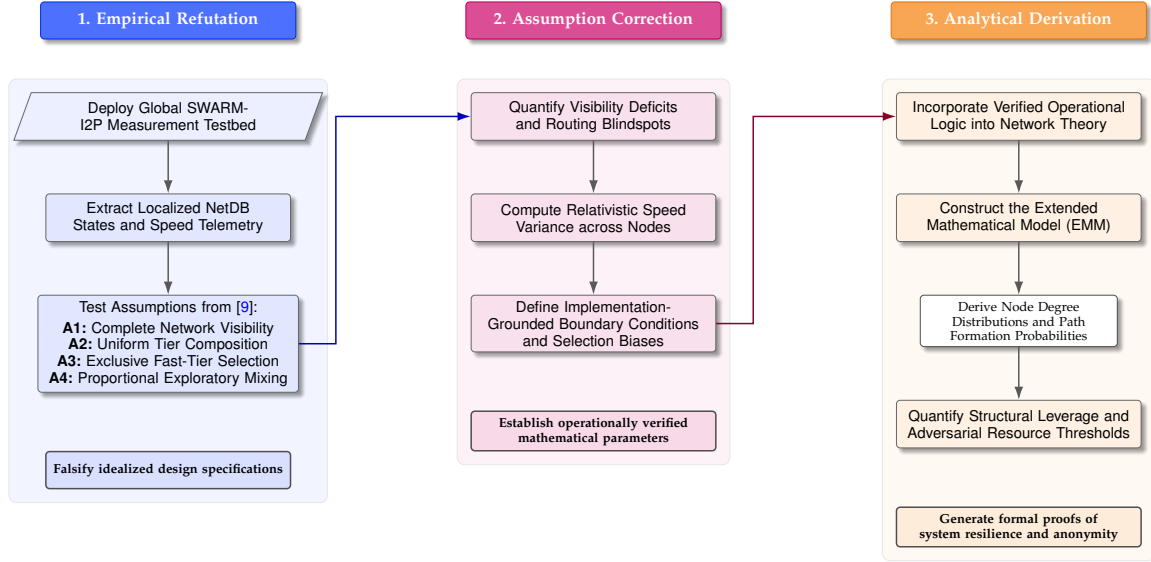
\begin{figure*}[htb]
\centering
\scalebox{0.70}{ 
\begin{tikzpicture}[
    node distance=1.0cm and 1.1cm,
    font=\footnotesize\sffamily,
    block/.style={rectangle, rounded corners=2pt, draw=black!60, line width=0.4pt, text width=4.4cm, minimum height=0.9cm, align=center, fill=#1!12, blur shadow={shadow blur steps=4}},
    io/.style={trapezium, trapezium left angle=70, trapezium right angle=110, draw=black!60, line width=0.4pt, text width=4.1cm, minimum height=0.9cm, align=center, fill=#1!8, blur shadow={shadow blur steps=4}},
    titlebox/.style={rectangle, fill=#1!70, draw=#1!90!black, text=white, font=\bfseries\small, minimum height=0.7cm, text width=4.6cm, align=center, rounded corners=2pt, blur shadow={shadow blur steps=5}},
    purpose/.style={rectangle, rounded corners=2pt, draw=black!70, thick, fill=#1!15, text width=4.4cm, align=center, font=\scriptsize\bfseries, minimum height=0.7cm},
    arrow_style/.style={-{Latex[length=2.5mm,width=1.7mm]}, thick, draw=black!65},
    phase_container/.style={rectangle, rounded corners=4pt, draw=black!10, fill=#1!5, inner sep=0.25cm},
    formula_block/.style={rectangle, draw=black!70, fill=white, text width=3.5cm, align=center, font=\scriptsize, rounded corners=2pt, line width=0.4pt, blur shadow={shadow blur steps=4}},
    title_text/.style={font=\bfseries\large\sffamily, align=center, text centered, text=black}
]

% ---- Title ----
\node[title_text, yshift=0.4cm] (overall_title) at (0,6.3)
{Methodological Framework: Empirical Refutation and Formal Derivation};

% ======================
% === Block 1 ==========
% ======================
\node[titlebox=blue!80!cyan] (p1title) at (-5,5.2) {1. Empirical Refutation};
\node[io=blue!80!cyan, below=of p1title] (src) {Deploy Global SWARM-I2P Measurement Testbed};
\node[block=blue!80!cyan, below=of src] (rules) {Extract Localized NetDB States and Speed Telemetry};
\node[block=blue!80!cyan, below=of rules, text width=4.7cm] (uml) {Test Assumptions from \cite{abdo2023modeling}:\\[2pt]
\textbf{A1:} Complete Network Visibility\\
\textbf{A2:} Uniform Tier Composition\\
\textbf{A3:} Exclusive Fast-Tier Selection\\
\textbf{A4:} Proportional Exploratory Mixing};
\node[purpose=blue!80!cyan, below=of uml, yshift=-0.1cm] (p1purpose) {Falsify idealized design specifications};

% ======================
% === Block 2 ==========
% ======================
\node[titlebox=purple!80!magenta, right=3.2cm of p1title] (p2title) {2. Assumption Correction};
\node[block=purple!80!magenta, below=of p2title] (model) {Quantify Visibility Deficits and Routing Blindspots};
\node[block=purple!80!magenta, below=of model] (predict) {Compute Relativistic Speed Variance across Nodes};
\node[block=purple!80!magenta, below=of predict] (hypotheses) {Define Implementation-Grounded Boundary Conditions and Selection Biases};
\node[purpose=purple!80!magenta, below=of hypotheses, yshift=-0.1cm] (p2purpose) {Establish operationally verified mathematical parameters};

% ======================
% === Block 3 ==========
% ======================
\node[titlebox=orange!85!brown, right=3.2cm of p2title] (p3title) {3. Analytical Derivation};
\node[block=orange!85!brown, below=of p3title] (testbed) {Incorporate Verified Operational Logic into Network Theory};
\node[block=orange!85!brown, below=of testbed] (collect) {Construct the Extended Mathematical Model (EMM)};
\node[formula_block, below=of collect] (formula) {Derive Node Degree Distributions and Path Formation Probabilities};
\node[block=orange!85!brown, below=of formula] (analyze) {Quantify Structural Leverage and Adversarial Resource Thresholds};
\node[purpose=orange!85!brown, below=of analyze, yshift=-0.1cm] (p3purpose) {Generate formal proofs of system resilience and anonymity};

% ---- Arrows ----
\foreach \i/\j in {src/rules, rules/uml} \draw[arrow_style] (\i) -- (\j);
\foreach \i/\j in {model/predict, predict/hypotheses} \draw[arrow_style] (\i) -- (\j);
\foreach \i/\j in {testbed/collect, collect/formula, formula/analyze} \draw[arrow_style] (\i) -- (\j);

% ---- Cross-phase Arrows ----
\draw[arrow_style, blue!70!black] (uml.east) -- ++(0.7,0) |- (model.west);
\draw[arrow_style, purple!70!black] (hypotheses.east) -- ++(0.7,0) |- (testbed.west);

% ---- Background Containers ----
\begin{scope}[on background layer]
    \node[phase_container=blue!80!cyan, fit=(src)(p1purpose)] {};
    \node[phase_container=purple!80!magenta, fit=(model)(p2purpose)] {};
    \node[phase_container=orange!85!brown, fit=(testbed)(p3purpose)] {};
\end{scope}

\end{tikzpicture}
} 
\caption{The sequential approach to model I2P's emergent topology. Step 1 uses a controlled testbed to empirically refute 
previous documentation-based 
assumptions~\cite{abdo2023modeling}. Step 2 quantifies these empirical divergences to derive 
corrected, implementation-grounded boundary conditions. Step 3 then uses these confirmed parameters to build the Extended Mathematical Model, rooting complex systems analysis squarely in operational reality.}
\label{fig:i2p_methodology_pipeline}
\end{figure*}

The methodology adopted in this study is a sequential methodology as illustrated in Fig.~\ref{fig:i2p_methodology_pipeline}. We begin with an empirical test, where we experiment with the major assumptions of past documentation-based models on a worldwide testbed of instrumented I2P routers. Second, after identifying the spheres of empirical divergence, we develop a new set of corrected, implementation-derived assumptions. These tested boundary conditions are later used to arrive at the Extended Mathematical Model (EMM) that gives the real topological properties of the network as an analytical consequence of its implementation-grounded peer-selection mechanisms.

\begin{figure}[htb]
\centering
\includegraphics[width=0.50\textwidth]{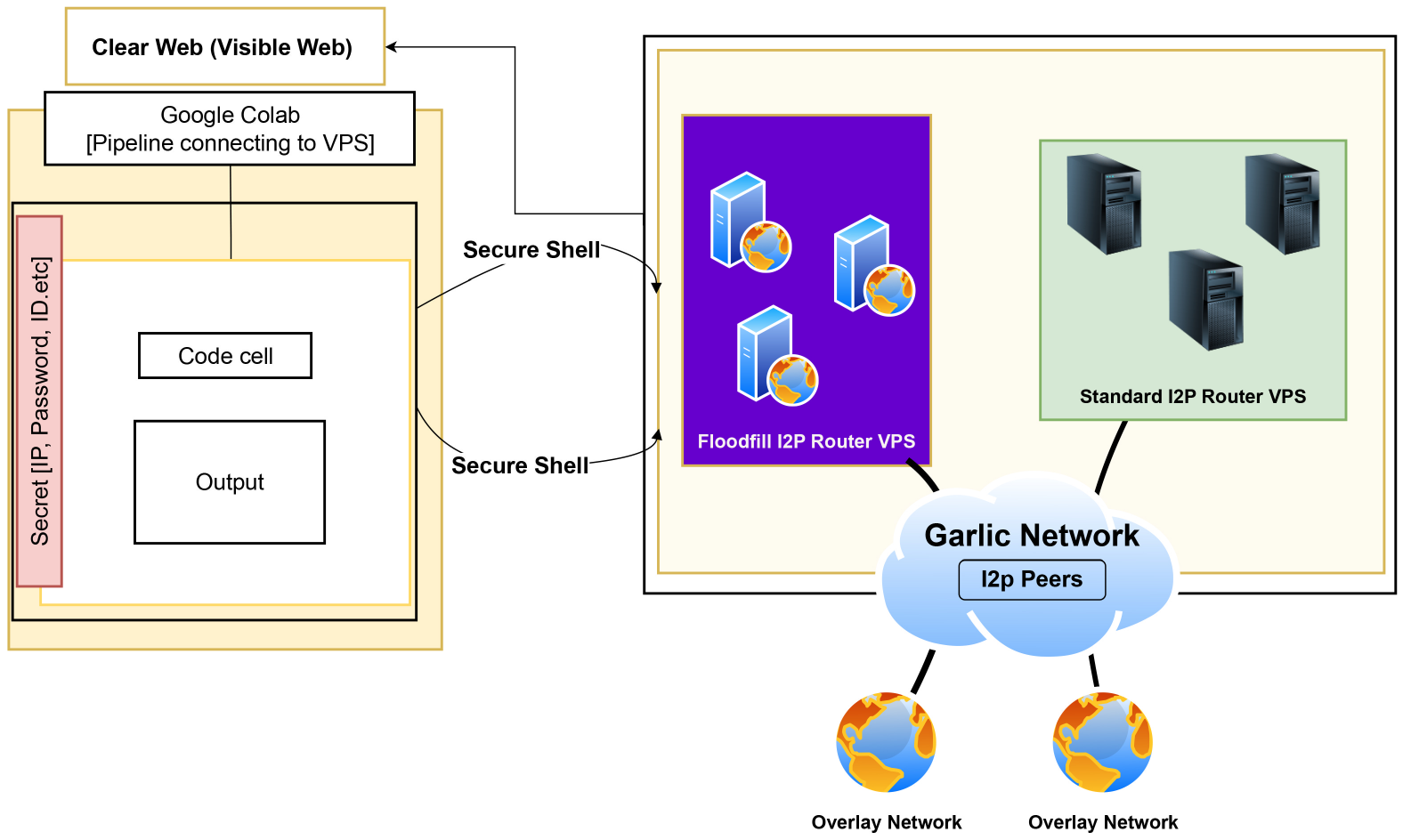}
\caption{Deployment architecture for the empirical validation study and global network measurement.}
\label{fig:cloudswarm}
\end{figure}

\subsection{Deployment of Testbed Experiments} \label{sec:deployment}

In order to facilitate the rigorous empirical measurement of peer selection mechanisms and network topology emergence, we implemented a controlled experimental infrastructure based on SWARM-I2P, as utilized by Muntaka et al.~\cite{muntaka2025mapping}. The testbed, as described in Fig.~\ref{fig:cloudswarm}, comprised six instances of I2P routers spread across geographically distributed locations using Virtual Private Servers (VPS). The routers were deployed natively into the active I2P Garlic Network and operated continuously for a minimum of thirty days per observation cycle. Our repository contains the deployment script, including floodfill configuration matrices and tunnel expansion parameters~\cite{siddique_i2pemm}.

The deployment architecture incorporated both floodfill and standard router configurations to reflect the operational heterogeneity of the live I2P network. Floodfill routers are the structural nodes with a very high capacity in which the network database (NetDB) is stored and consulted by other routers (nodes) during tunnel construction. Each VPS was set up with the I2P router version 2.10.0 on Ubuntu 24.04 LTS and executed automation scripts~\cite{siddique_i2pemm} to 
systematically vary tunnel configuration parameters across 
observation cycles.

The router profile states (via the \texttt{/profiles} endpoint) and active tunnel peer assignments (via the \texttt{/tunnels} endpoint) had to be extracted at synchronized timestamps. This synchronization over time ensured that the state of the observed local network and the peer selections matched. This dual-dataset design enables direct mathematical correlation between the available peer pools captured in profile snapshots and the selection patterns recorded in tunnel compositions, a capability absent from prior measurement studies that examined profiles and tunnel selections in isolation.

The obtained datasets are recorded in our GitHub repository~\cite{siddique_i2pemm} as notebooks reflecting the assumptions of the first novel analytical bottom-up studies of I2P~\cite{abdo2023modeling}:

\begin{enumerate}
    \item \texttt{1\_Routers\_have\_FullView.ipynb}: Evaluation of network visibility constraints.
    \item \texttt{2\_Routers\_have\_similar\_View.ipynb}: Comparative analysis of NetDB consistency across disparate geographic locations.
    \item \texttt{3\_ClientTunnelSelectFastset.ipynb}: Empirical measurement of client tunnel peer selection patterns.
    \item \texttt{4\_ExploratoryTunnelSelectstandard\_High\\
    capa.ipynb}: Characterization of exploratory tunnel tier preferences.
    \item \texttt{5\_RouterProfileView\_All.ipynb}: Comprehensive profiling of peer tier distributions.
\end{enumerate}

These datasets constitute the empirical basis for our examination of the documentation-based modeling assumptions and the development of our calibrated analytical model.

\subsection{Empirical Refutation of Prior Analytical Assumptions} \label{sec:critique}

The seminal bottom-up analytical model for I2P's emergent topology developed by Bou Abdo and Hossain~\cite{abdo2023modeling} represented a critical advancement toward formal analytical prediction. However, their model relied on four specific assumptions about operational behavior derived directly from interpreting protocol specifications. Through controlled testbed experiments, we demonstrate that these assumptions 
systematically diverge from the operational behavior 
of the deployed network. 

\textbf{Assumption 1 - Complete Network Visibility:} The model in~\cite{abdo2023modeling} presumes that individual routers command a complete picture of the network. This is formally stated as $\Omega = \Omega_s$, where $\Omega_s$ is the set of peers visible to router $s$, and the total number of peers on the network as $\Omega$. This assumption implies that every member has a homogeneous knowledge of the network space such that peer selection decisions become universal.

\begin{figure}[H]
\centering
\includegraphics[width=0.49\textwidth]{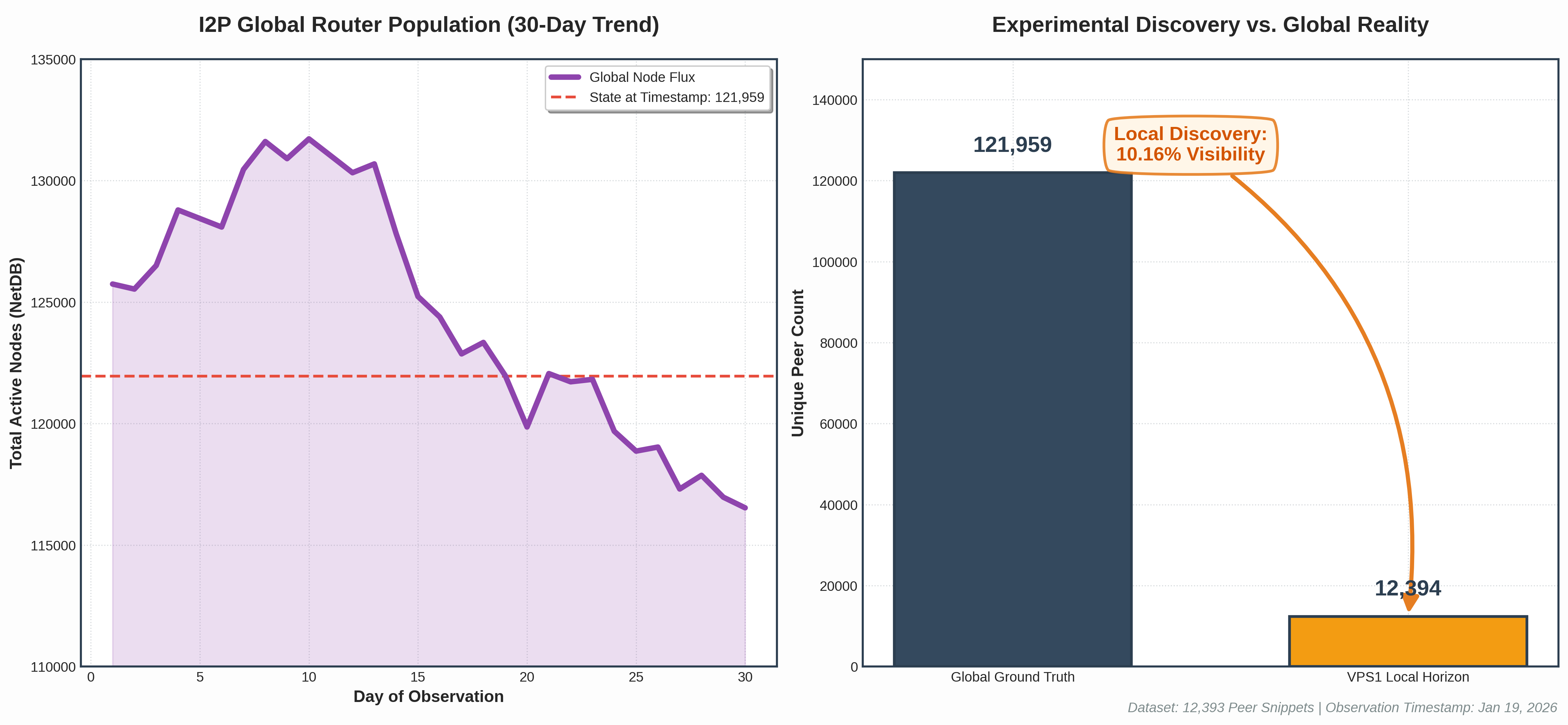}
\caption{The partial network visibility in I2P is illustrated. Left: Fluctuation of the global router population displaying 121,959 unique peers. Right: VPS1's local NetDB had 12,394 peers, which is 10.16\% visibility. This empirical measurement directly refutes the assumption of complete network visibility in documentation-based models, namely that $\Omega = \Omega_s$.}
\label{fig:network_visibility}
\end{figure}

\textbf{Empirical Refutation:} The extraction of localized states of the NetDB indicated a highly heterogeneous visibility of the network. The local NetDB of any single router captures only a small fraction of the total network population. The current consensus statistics (according to stats.i2p) indicate that there are about 121,959 active peer identities in the network over a 30-day period (see Fig.~\ref{fig:network_visibility}). A baseline of around 50,000 active routing nodes was established by systematic enumeration in Muntaka et al.~\cite{muntaka2025mapping}. It is against this structural background that our experiments confirmed that individual routers operate 
with severely constrained network visibility. VPS1, having operated continuously during the 30 days, only had 12,394 peers in its NetDB. This represents low visibility of the global network state (10.16\%), and approximately 24.8\% visibility of the active routing nodes.

This constraint imposes an analytical boundary which ultimately defines the emergent topology of the network. In the construction of tunnels by a router, it mathematically eliminates all but a small fraction of the network. A peer that has the best routing capacity is still given a selection weight of zero in the event that they do not happen to be in the localized view of the executing router. The emergent topology is thus not based on a globally consistent pool of peers, but rather on an aggregated distribution of more constrained, fragmented perspectives of the network. Models that do not capture this structural blindness lead to flawed security assumptions due to the computation of anonymity guarantees over a hypothetical denominator of a complete network that none of the routers in reality reaches.

\begin{figure}[H]
\centering
\includegraphics[width=0.48\textwidth]{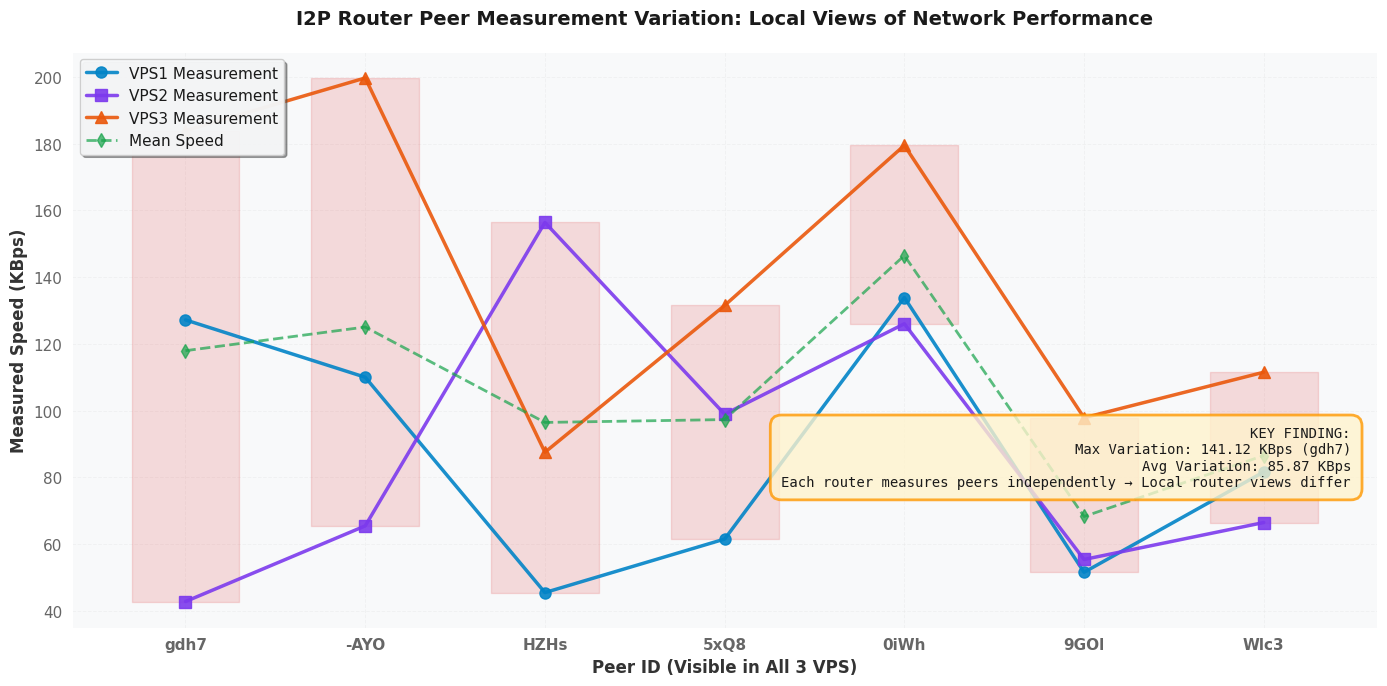}
\caption{Heterogeneity in speed measurement among three I2P routers for identical peers. The shaded areas indicate substantial variance (up to 141.12 KBps difference when comparing a single peer across routers), confirming that tier assignment is relative to the network position of the observing router.}
\label{fig:speed_variation}
\end{figure}

\textbf{Assumption 2 - Homogeneous Level of Peer Tier Composition by Router:} In the model of Bou Abdo and Hossain~\cite{abdo2023modeling}, all routers develop the performance tiers (Fast, High Capacity, Standard) homogeneously. In theory, the set of peers router $s_1$ classifies as Fast, the set of peers router $s_2$ classifies as Fast, and the set of peers router $s_3$ classifies as Fast are identical, i.e., $\mathcal{F}_{s_1} = \mathcal{F}_{s_2} = \mathcal{F}_{s_3}$. This is based on the assumption that peer profiling results in absolute and globally consistent classifications.

\textbf{Empirical Refutation:} A synchronized controlled experiment on three testbed routers, located in geographically distant areas, demonstrated that peer classification showed significant disagreement. The test assessed 450 consensus peers at the same time. Only 7 peers (1.6\%) achieved uniform classification as Fast or High Capacity across all three observational vantage points. A further 79 peers (17.6\%) achieved consensus between two routers, while 271 peers (60.2\%) were categorized into high-performance tiers by only a single router.

Fig.~\ref{fig:speed_variation} shows the mechanism driving this divergence. Network path disparities and geographic routing latency guarantee that router $s_1$'s measurement of peer $p$ will fundamentally differ from router $s_2$'s measurement. The recorded speed for identical peers varied drastically. For instance, peer \texttt{gdh7} registered at 127.17 KBps for Router 1, 42.77 KBps for Router 2, and 183.89 KBps for Router 3, resulting in a variation of 141.12 KBps. The mean variation spread across all seven consensus peers reached 85.87 KBps. Because tier boundaries are derived locally by ranking these divergent speed measurements, each router generates entirely unique tier populations. Therefore, $\mathcal{F}_{s_1} \neq \mathcal{F}_{s_2}$. Theoretical structures that assume a homogeneous tier structure fail to reflect the local, relativistic mathematical realities that govern the chances of forming tunnels.

\textbf{Assumption 3 - Selecting Client Tunnels Exclusively from the Fast Tier:} The model in~\cite{abdo2023modeling} makes a strict assumption that client tunnels are constructed based on selection probabilities restricted exclusively to the Fast tier.

\begin{figure}[htb]
\centering
\includegraphics[width=0.50\textwidth]{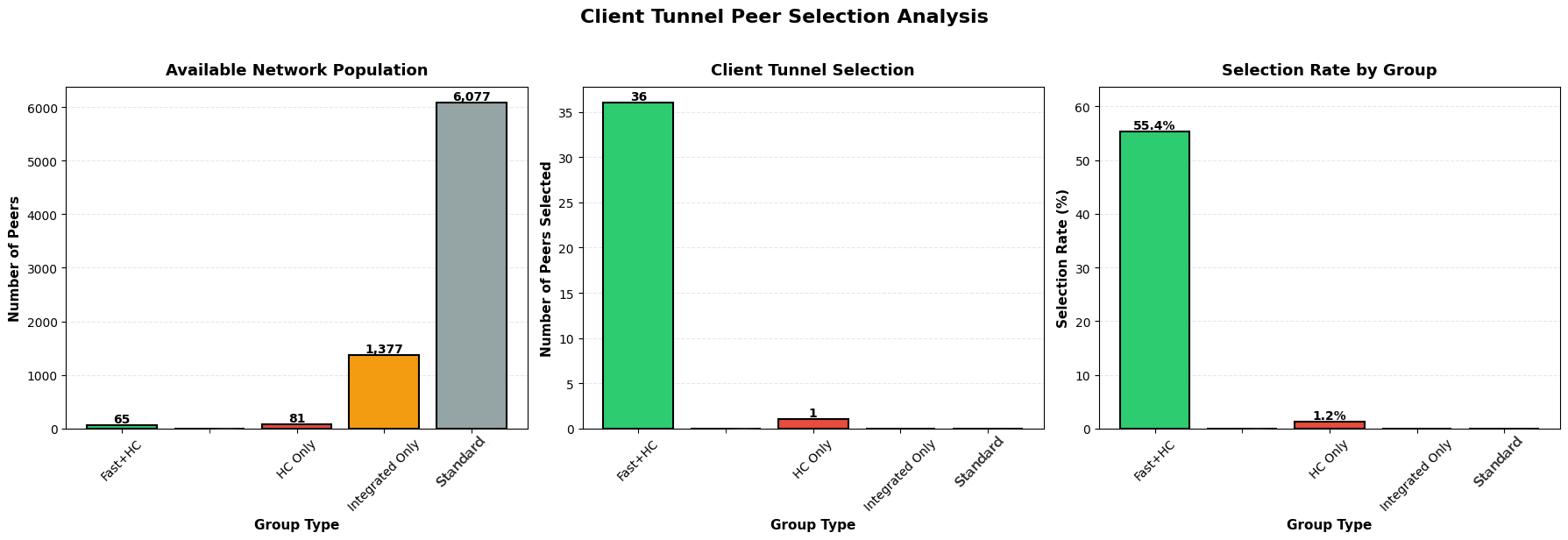}
\caption{Client tunnel peer selection patterns. The right panel measures a 55.4\% selection rate for available Fast tier peers but confirms a non-zero (1.2\%) opportunistic fallback selection from the High Capacity tier, refuting the assumption of absolute Fast-tier exclusivity.}
\label{fig:client_tunnel}
\end{figure}

\textbf{Empirical Refutation:} Although empirical evidence shows a heavy weighting toward Fast tier nodes, the choice is mathematically non-exclusive. Fig.~\ref{fig:client_tunnel} describes one such operational sample in which a router's database had 65 Fast/High-Capacity peers and 81 High-Capacity-only peers. In the case of its client tunnels, the router chose 37 peers total: 36 from the Fast tier (97.3\%) and 1 from the High-Capacity-only pool (2.7\%). 

This confirms an operationally significant behavior absent 
from idealized models. The protocol performs an opportunistic fallback to the broader High Capacity pool when the Fast tier is momentarily depopulated by natural node churn. Since the tier profile is dynamically rearranged according to real-time performance medians, this fallback probability is mathematically continuous. An implicit assumption of absolute Fast tier exclusivity essentially constrains the analytical modelling of adversarial node injection strategies during these specific fallback windows.

\begin{figure}[htb]
\centering
\includegraphics[width=0.50\textwidth]{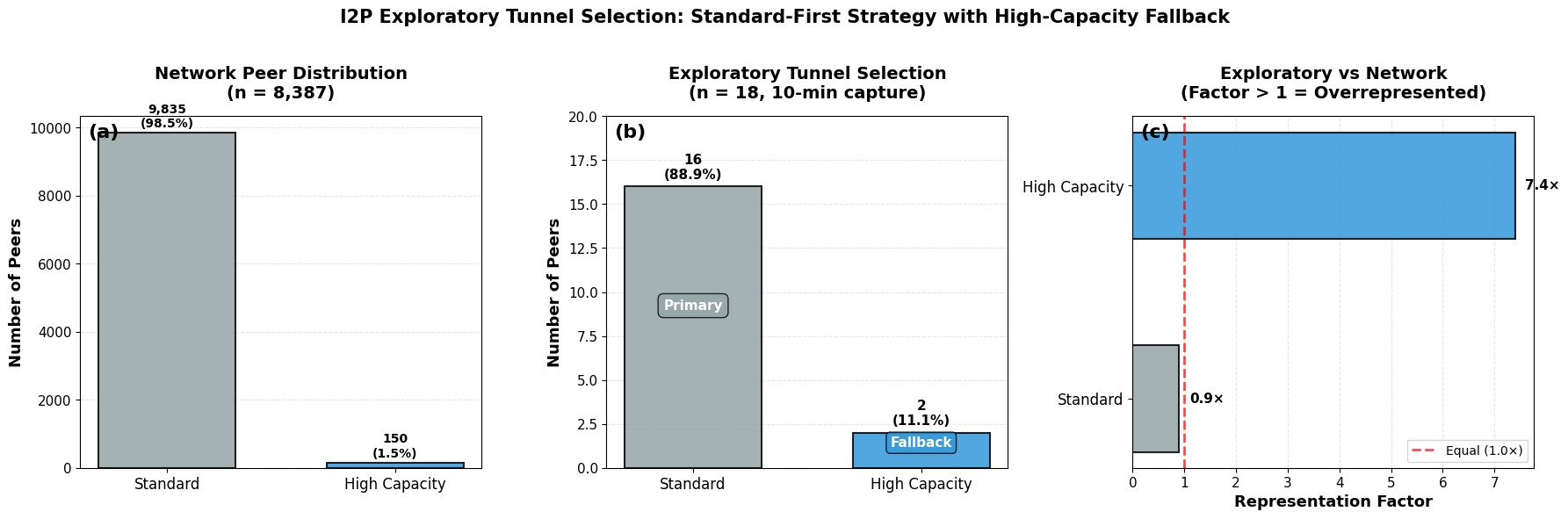}
\caption{Exploratory tunnel peer selection demonstrating a 7.4$\times$ overrepresentation of High Capacity peers. This indicates deliberate algorithmic preference, refuting models that assume strictly proportional tier sampling.}
\label{fig:exploratory_tunnel}
\end{figure}

\textbf{Assumption 4 - Proportional Selection for Exploratory Tunnels:} Assumptions in prior modelling~\cite{abdo2023modeling} dictate that exploratory tunnels pick peers based on a fixed proportional mixture of the Standard and High Capacity tiers, formulated as $\alpha S + (1-\alpha)HC$, which is directly proportional to the network population.

\textbf{Empirical Refutation:} The existence of a tier mixture is confirmed by investigations of exploratory tunnel telemetry, but reveals a deep quantitative discrepancy with homogeneous proportional mixing. As seen in Fig.~\ref{fig:exploratory_tunnel}, the network database had 9,835 Standard peers (98.5\%) and 150 High Capacity peers (1.5\%). Still, during a normalized observation capturing 18 peer selections, the router chose 16 Standard peers (88.9\%) and 2 High Capacity peers (11.1\%). 

Calculation of the representation factor reveals a deliberate algorithmic bias. High Capacity peers are overrepresented by 7.4$\times$ relative to their proportion in the network population. Mathematically, this proves that uniform sampling during exploratory selection does not occur. It implements a Standard-first policy coupled with an aggressive, non-proportional High Capacity choice bias to provide minimum performance guarantees during network discovery. The empirically observed mixing proportion ($\alpha \approx 0.889$) diverges substantially from the proportional assumption ($\alpha \approx 0.985$).

\subsection{Proposing Corrected Assumptions for the Extended Mathematical Model} \label{sec:realistic}

Since the empirical disjuncture between idealized documentation and operational reality has been demonstrated, parameters are needed that describe the actual behavior of the network as it is implemented. Through the combination of empirical boundary conditions that we found in our testbed deployment, we have been able to derive a recalibrated set of assumptions which are based on real implementation. These updated postulates provide the tested mathematical basis needed to build the Extended Mathematical Model, allowing our analysis to transition from testing earlier models to formally deriving I2P's emergent topology.

\section{Extended Mathematical Model} \label{EMM}

This section extends the mathematical model of \cite{abdo2023modeling} under the corrected, implementation-validated assumptions established in 
Section~\ref{sec:realistic}, retaining the original nomenclature where applicable.

% SID ADD a subsection on data collection for validation\\ - Kindly vert I added after the model

% ========== INSERT ALGORITHM HERE ==========

Having established through empirical validation (Section~\ref{sec:critique}) that documentation-based assumptions systematically diverge from operational behavior, we now formalize the implementation-grounded peer selection mechanism that drives I2P's emergent topology. Algorithm~\ref{alg:tier_classification} presents the tier classification procedure, capturing how individual routers transform locally observed peer performance measurements into the stratified structure underlying our Extended Mathematical Model.

\begin{algorithm}[t]
\caption{Local Router Peer Profiling and Tier Classification}
\label{alg:tier_classification}
\begin{algorithmic}[1]
\small
\State \textbf{Input:}
\State \quad $\omega_s$: Set of active, non-failing peers in router $n_s$'s NetDB
\State \quad $\mathbf{M}_{\text{speed}}$: Measured speed values for peers in $\omega_s$
\State \quad $\mathbf{M}_{\text{capacity}}$: Measured capacity values for peers
\State \textbf{Output:}
\State \quad $\gamma_s$: Fast tier peers
\State \quad $g_{1s}$: High Capacity tier peers
\State $\tau_{\text{capacity}} \gets \text{mean}(\{\mathbf{M}_{\text{capacity}}[n_i] : n_i \in \omega_s\})$
\State Sort $\omega_s$ by capacity (highest first)
\State $\mathcal{HC}_{\text{prelim}} \gets \{n_i \in \omega_s : \mathbf{M}_{\text{capacity}}[n_i] \geq \tau_{\text{capacity}}\}$
\State $\tau_{\text{speed}} \gets \text{mean}(\{\mathbf{M}_{\text{speed}}[n_i] : n_i \in \mathcal{HC}_{\text{prelim}}\})$
\State $g_{1s} \gets \emptyset$, $\gamma_s \gets \emptyset$
\For{each peer $n_i \in \omega_s$}
    \If{$\mathbf{M}_{\text{capacity}}[n_i] \geq \tau_{\text{capacity}}$ \textbf{and} $\text{isSelectable}(n_i)$}
        \State $g_{1s} \gets g_{1s} \cup \{n_i\}$
        \If{$\mathbf{M}_{\text{speed}}[n_i] \geq \tau_{\text{speed}}$ \textbf{and} $\text{isActive}(n_i)$}
            \State $\gamma_s \gets \gamma_s \cup \{n_i\}$
        \EndIf
    \EndIf
\EndFor
\If{$|\gamma_s| > 75$}
    \State Sort $\gamma_s$ by speed (slowest first)
    \State Remove slowest $(|\gamma_s| - 75)$ peers from $\gamma_s$
\EndIf
\If{$|g_{1s}| > 150$}
    \State Sort $g_{1s}$ by capacity (lowest first)
    \State Remove lowest $(|g_{1s}| - 150)$ peers from $g_{1s}$
    \State Also remove these peers from $\gamma_s$ if present
\EndIf
\State \Return $\gamma_s$, $g_{1s}$
\end{algorithmic}
\end{algorithm}

This algorithmic formalization reveals three critical implementation details absent from documentation-based models. First, tier membership is determined by threshold comparison against mean values, not percentile ranking. Second, the Fast tier is nested within the High Capacity tier ($\gamma_s \subseteq g_{1s}$), as Fast peers must first satisfy capacity thresholds. Third, hard size limits (75 Fast, 150 High Capacity) are enforced through demotion of marginal peers, creating the bounded tier populations observed in our empirical validation (Section~\ref{sec:critique}).

The mathematical formalization that follows employs this algorithmic logic to derive expected router degree distributions under partial network visibility and heterogeneous local profiling.

Considering four complete homogeneous graphs $\Gamma$, $G_{1}$,  $G_{2}$ and $\Omega$ of sizes $N_{\gamma}$, $N_{g1}$, $N_{g2}$ and $N_{\omega}$ respectively, where $N_{\gamma} > 2$, $N_{g1}>2$,  $N_{g2}>2$ and $N_{\omega}>2$. Let $\gamma$, $g_{1}$, $g_{2}$ and $\omega$ be the set of nodes in $\Gamma$, $G_{1}$, $G_{2}$ and $\Omega$ respectively, defined as $\gamma= \{n_{1}, n_{2}, ..., n_{N\gamma}\}$, $g_{1}= \{n_{1}, n_{2}, ..., n_{N_{g1}}\}$, $g_{2}= \{n_{1}, n_{2}, ..., n_{N_{g2}}\}$ and $\omega = \{n_{1}, n_{2}, ..., n_{N\omega}\}$. Let $\gamma \subset G_{1}  \subset G_{2} \subset \omega$ and thus $\Gamma$ a subgraph of $G_{1}$, which is a subgraph of $G_{2}$, which is a subgraph of $\Omega$. 

The sets $\gamma_s$, $g_{1s}$, $g_{2s}$ and $\omega_s$ represent the Fast, High Capacity, Standard and RouterInfo sets observed by router $n_{s}$ respectively, Algorithm~\ref{alg:tier_classification} formalizes the tier classification procedure that partitions $\omega_s$ into these 
performance-based strata. The set $\omega_s$, viewed by router $n_{s}$, is a subset of the space of routers constituting the I2P network, denoted $\omega_{I2P}$ (with size $N_{\omega_{I2P}}$).

Let $T$ be a client tunnel, which is a linear chain graph, starting with node $n_{s}$. Let $\tau$ be the set of nodes in $T$. Since $n_{s}$ attempts to establish its client tunnel $T_s$ from the set of Fast routers within its view (i.e. $\tau_s \subset \gamma_s$), but when it fails, it attempts to establish its client tunnel $T_s$ from the set of High Capacity routers within its view (i.e. $\tau_s \subset g_{1s}$), then let $\alpha$ be the probability of $n_{s}$ successfully selecting the client tunnel $T_s$ from the set of Fast routers within its view.

\begin{corollary}\label{corollary1} 
The probability of a node $n_i$ to be selected in a client tunnel initiated by $n_s$, knowing that $n_i$ is not in $n_s$'s view is 0:\\
\begin{equation}\label{eq1}
    P(n_{i} \in  \tau_s \ | \ n_{i} \notin \omega_s) = 0
\end{equation}
According to the \cite{muntaka2026systemic} and corrected assumption 1 ($\omega_s \subset \omega_{I2P}$).
\end{corollary}

\begin{corollary}\label{corollary2} 
The probability of a node $n_i$ to be selected in an exploratory tunnel initiated by $n_s$, knowing that $n_i$ is not in $n_s$'s view is 0:\\
\begin{equation}\label{eq1.1}
    P(n_{i} \in  \tau_{es} \ | \ n_{i} \notin \omega_s) = 0
\end{equation}
According to the \cite{muntaka2026systemic} and corrected assumption 1 ($\omega_s \subset \omega_{I2P}$).
\end{corollary}

\begin{corollary}\label{corollary3} 
The probability of a node $n_i$ to be in $n_s$'s view is:\\
\begin{equation}\label{eq2}
    P(n_{i} \in \omega_s) = \frac{N_{\omega_{s}}}{N_{\omega_{I2P}}}
\end{equation}
According to \cite{muntaka2026systemic} corrected assumption 1 ($\omega_s \subset \omega_{I2P}$).
\end{corollary}

As demonstrated in Fig.~\ref{fig:network_visibility} and section \ref{sec:critique}, a router's network view is around $25\%$ of the total population of peers ($12,394$ peers out of $50,000$). This results in the Cumulative Distribution Function (CDF) of $\omega_s$ being an empirical estimator of $\omega_{I2P}$'s CDF.

\begin{corollary}\label{corollary4}
Node $n_i$'s CDF is:\\
\begin{equation}\label{eq2.5}
    CDF_s(n_{i}) \approx \begin{cases}
 CDF_{I2P}(n_{i}), & if \ n_i \in \omega_s\\
0, & otherwise\\
\end{cases}
\end{equation}
\end{corollary}

\begin{corollary}\label{corollary5} 
The probability of a node $n_{i}$ to be part of $\tau_s$, knowing that $n_{i} \in \gamma_s$ is:\\
\begin{equation}\label{eq3}
    P(n_{i} \in  \tau_s \ | \ n_{i} \in \gamma_s) = \frac{H \times N_{\omega_s} + N_{\gamma_s}}{N_{\omega_s} \times N_{\gamma_s}}
\end{equation}
According to eq. 2 of \cite{abdo2023modeling}. To validate the correctness of this corollary, we compared the observed frequency a node is listed in the client tunnel against equation \ref{eq3}, as shown in "corollary 5" tab in  \cite{siddique_i2pemm} (excel file in Github "data" repository). We used Chi-square two sample test (observed vs. the predicted according to equation \ref{eq3}). The p-value scored $0.976$ indicating that there is not a significant difference between the two samples. Accordingly, this validates equation \ref{eq3}. 
\end{corollary}

\begin{corollary}\label{corollary6} 
The probability of a node $n_{i}$ to be in the fast set, knowing that it is in the high capacity set is:\\
\begin{equation}\label{eq4}
    P(n_{i} \in \gamma_s \ | \ n_{i} \in N_{g1s})   \qquad \qquad \qquad \qquad \qquad \qquad \qquad \qquad \\ 
\end{equation}
\[=\begin{cases}
0 & if \ CDF(n_i) < 0.5\\
0 & if \ CDF(n_{(N_{g1s}-MFS)}) > CDF(n_i) \ge 0.5\\
1 & otherwise\\
\end{cases}\]
where $CDF(n_i)$ represents $CDF_{s}(n_i)$, but can be replaced by $CDF_{I2P}(n_i)$ with appropriate index adjustment, according to equation \ref{eq2.5}. $CDF$ is over the speed distribution as observed by $n_s$. $MFS$ represents the Maximum Fast Set constant (currently $75$ as indicated in Fig.~\ref{fig:i2p_arch_profile} right) determined in I2P's source code \cite{muntaka2026systemic}. To validate the correctness of this corollary, we compared the observed frequency a node is listed in the fast set against equation \ref{eq4}, as shown in "corollary 6" tab in \cite{siddique_i2pemm}(excel file in Github "data" repository). We used Chi-square two sample test (observed vs. the predicted according to equation \ref{eq4}). The p-value scored $1$ indicating perfect match between the two samples. Accordingly, this validates equation \ref{eq4}.
\end{corollary}

\begin{corollary}\label{corollary7} 
The probability of a node $n_{i}$ to be in the high capacity set, knowing that it is within the view of $n_s$ is:\\
\begin{equation}\label{eq5}
    P(n_{i} \in N_{g1s} \ | \ n_{i} \in \omega_s)   \qquad \qquad \qquad \qquad \qquad \qquad \qquad \qquad \\ \end{equation}
\[
=\begin{cases}
0 & if \ CDF(n_i) < 0.5\\
0 & if \ CDF(n_{(N_{\omega_s}-MHCS)}) > CDF(n_i) \ge 0.5\\
1 & otherwise\\
\end{cases}\]
where $CDF(n_i)$ represents $CDF_{s}(n_i)$, but can be replaced by $CDF_{I2P}(n_i)$ with appropriate index adjustment, according to equation \ref{eq2.5}. $CDF$ is over the capacity distribution as observed by $n_s$. $MHCS$ represents the Maximum High Capacity Set constant (currently $150$) determined in I2P's source code \cite{muntaka2026systemic}. To validate the correctness of this corollary, we compared the observed frequency a node is listed in the fast set against equation \ref{eq5}, as shown in "corollary 7" tab in \cite{siddique_i2pemm} (excel file in Github "data" repository). We used the Kolmogorov–Smirnov (K–S) test (observed vs. the predicted according to equation \ref{eq5}), since both samples contain zeros. The KS statistic scored $0$ indicating perfect match between the two samples. Accordingly, this validates equation \ref{eq5}.
\end{corollary}

\subsection{Derivation}

\begin{lemma}  \label{lemma1}
The probability of a node $n_{i}$ to be part of $\tau_s$ is:\\
\begin{equation}\label{eq2}
     P(n_{i} \in  \tau_s) =\begin{cases}
0 & if \ condition\\
\frac{H \times N_{\omega_s} + N_{\gamma_s}}{N_{I2P} \times N_{\gamma_s}} & otherwise\\
\end{cases}
\end{equation}
condition: $(CDF(n_{(N_{g1s}-MFS)}) > CDF(n_i) \ge 0.5) \ OR $\\
$(CDF(n_{(N_{\omega_s}-MHCS)}) > CDF(n_i) \ge 0.5 \ OR $\\
$(CDF_{speed}(n_i) < 0.5) \ OR \ (CDF_{capacity}(n_i) < 0.5) \ OR$\\
Derivation:
\[
 P(n_{i} \in  \tau_s) = P(n_{i} \in  \tau_s \ | \ n_{i} \in \gamma_s) \times P(\ n_{i} \in \gamma_s) \qquad \qquad \quad 
\]
\[
 = P(n_{i} \in  \tau_s \ | \ n_{i} \in \gamma_s) \times P(\ n_{i} \in \gamma_s \ | \ n_{i} \in g_{1s}) \times P(\ n_{i} \in g_{1s})
\]
\[
 = P(n_{i} \in  \tau_s \ | \ n_{i} \in \gamma_s) \times P(\ n_{i} \in \gamma_s \ | \ n_{i} \in g_{1s}) \times \qquad \qquad \quad 
\]
\[
 \qquad  \qquad P(\ n_{i} \in g_{1s} \ | \ n_{i} \in \omega_s) \times P(\ n_{i} \in\omega_s)
\]
\[
 = \frac{H \times N_{\omega_s} + N_{\gamma_s}}{N_{\omega_s} \times N_{\gamma_s}} \times \frac{N_{\omega_{s}}}{N_{\omega_{I2P}}} \times P(\ n_{i} \in \gamma_s \ | \ n_{i} \in g_{1s}) \times \qquad \qquad \quad 
\]
\[ \qquad \qquad \quad \qquad \qquad \qquad \quad P(\ n_{i} \in g_{1s} \ | \ n_{i} \in \omega_s) 
\]
\[
 =\begin{cases}
0 & if \ CDF_{speed}(n_i) < 0.5\\
0 & if \ CDF_{capacity}(n_i) < 0.5\\
0 &\!\!\!\!\!\!\!\!\!\!\!\!\!\!\! if \ CDF(n_{(N_{g1s}-MFS)}) > CDF(n_i) \ge 0.5\\
0 &\!\!\!\!\!\!\!\!\!\!\!\!\!\!\! if \ CDF(n_{(N_{\omega_s}-MHCS)}) > CDF(n_i) \ge 0.5\\
\frac{H \times N_{\omega_s} + N_{\gamma_s}}{N_{I2P} \times N_{\gamma_s}} & otherwise\\
\end{cases}
\]
\end{lemma}

Let $T_{es}$ be an exploratory tunnel, which is a linear chain graph, starting with node $n_{s}$. Let $\tau_{es}$ be the set of nodes in $T_{es}$. Since $n_{s}$ attempts to establish its exploratory tunnel $T_{es}$ from the set of Standard routers within its view (i.e. $\tau_{es} \subset g_{2s}$), but when it fails, it attempts to establish its exploratory tunnel $T_{es}$ from the set of High Capacity routers within its view (i.e. $\tau_s \subset g_{1s}$),

then let $\beta$ be the overrepresentation factor of High Capacity peers in exploratory tunnel selection relative to their network 
proportion, capturing the degree to which $n_{s}$ preferentially 
selects High Capacity peers beyond what uniform sampling would 
predict. $\beta$ is empirically measured and reported in Fig.~\ref{fig:exploratory_tunnel} as $7.4$, meaning High Capacity 
peers are selected at $7.4\times$ their network-proportional rate. Similar to lemma \ref{lemma1}, the probability of a node to participate in an exploratory tunnel is formulated in lemma \ref{lemma2}.

\begin{lemma} \label{lemma2}
The probability of a node $n_{i}$ to be part of $\tau_{es}$ is:\\
\begin{equation}\label{eq2}
     P(n_{i} \in  \tau_{es}) = \frac{H+1}{N_{I2P}} \times \qquad \qquad \qquad \qquad \qquad \qquad \qquad \qquad
 \end{equation}
 \[
     \begin{cases}
1 & if \ CDF(n_i) < 0.5\\
1 & if \ CDF(n_{(N_{\omega_s}-MHCS)}) > CDF(n_i) \ge 0.5\\
\beta & otherwise\\
\end{cases}
\]
\end{lemma}

\begin{theorem} \label{theorem_structure_4}
The degree of a router is expected to be:
\begin{equation}
     Deg(n_{i}) =  2 \times
     \end{equation}
    \[
{\small
\begin{cases}
(H+1) &\!\!\!\! if \ CDF_{speed}(n_i) < 0.5\\
(H+1) &\!\!\!\! if \ CDF_{capacity}(n_i) < 0.5\\
(H+1) &\!\!\!\!\!\!\!\!\!\!\!\!\! if \ CDF(n_{(N_{g1s}-MFS)}) > CDF(n_i) \ge 0.5\\
(H+1) &\!\!\!\!\!\!\!\!\!\!\!\!\!\!\! if \ CDF(n_{(N_{\omega_s}-MHCS)}) > CDF(n_i) \ge 0.5\\
\frac{H N_{\omega_s}}{N_{\gamma_s}} +1 + \beta (H+1) & otherwise\\
\end{cases}
}
\]

following section 5 in \cite{abdo2023modeling},\\
$Deg(n_{i}) = 2 \times N_{I2P} \times (P(n_{i} \in  \tau_s) + P(n_{i} \in  \tau_{es}))$
\end{theorem}

% Inser Map for Pedicted Emerging Structure

\begin{figure*}[htb]
\centering
\includegraphics[width=0.50\textwidth]{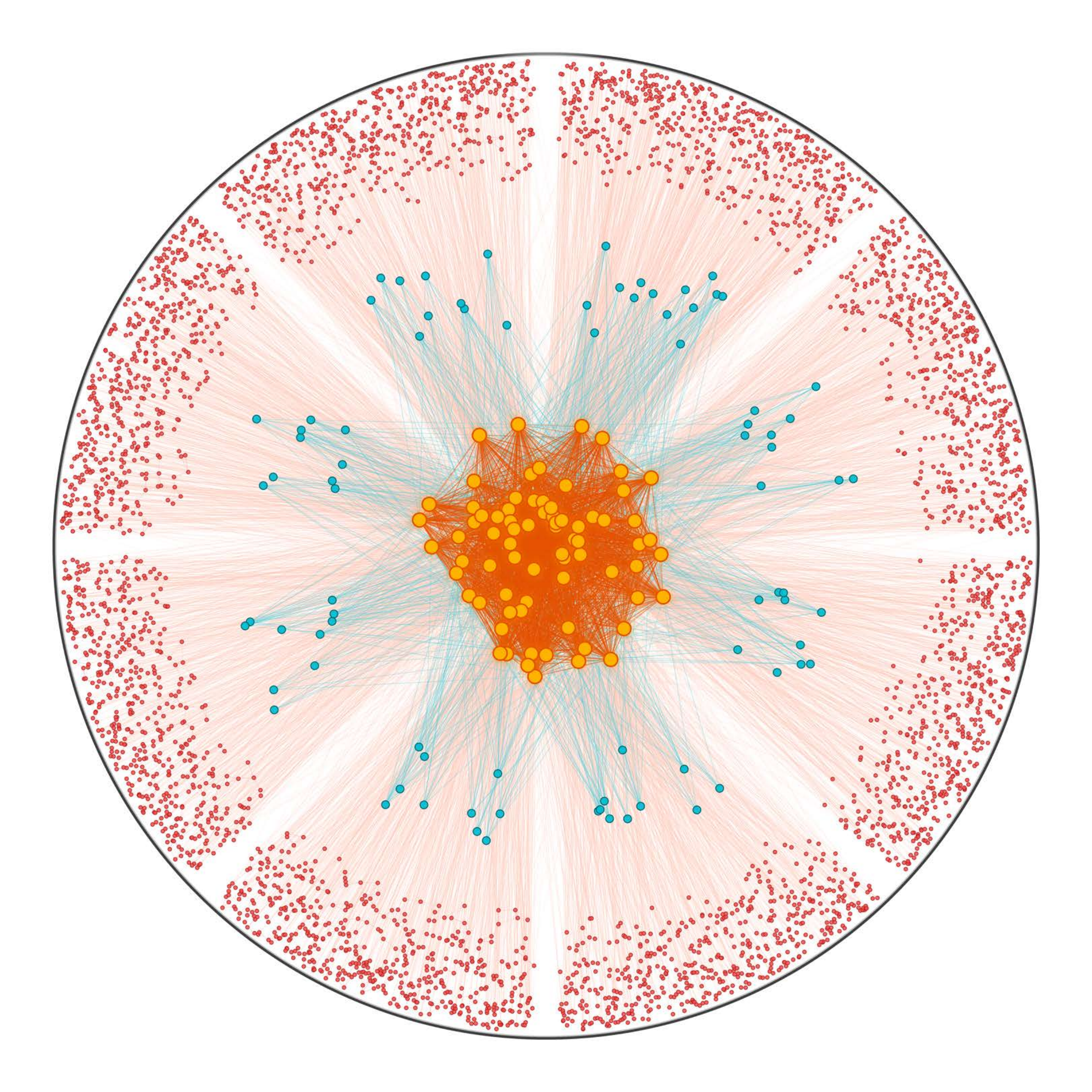}
\caption{EMM-predicted emergent topology of the I2P Garlic Network ($N_{\omega_{I2P}} \approx 50{,}000$), rendered on a Poincaré disk (model-predicted, not empirically measured). Node size encodes tier membership: Fast ($N_{\gamma_s} \leq \mathit{MFS} = 75$), High Capacity ($N_{g_{1s}} \leq \mathit{MHCS} = 150$), and Standard; colour encodes tier membership: Fast nodes (amber/gold), High Capacity nodes (cyan), and Standard nodes (coral); arc sectors are uniform partitions of the disk. Edges represent predicted tunnel connections (Theorem~\ref{theorem_structure_4}). Fast-tier hubs dominate the interior web with predicted degree ${\approx}1053$, exceeding Standard routers ($\mathit{Deg} = 8$) by $131.6\times$ despite their bounded population. Degree asymmetry across the disk is an emergent consequence of tier-stratified peer selection (Theorem~\ref{theorem_structure_4}), not a modeling artifact. No individual router observes this global structure; each operates within a local ${\approx}24.8\%$ network slice (Corollary~\ref{corollary3}).}
\label{fig:emm_topology}
\end{figure*}

% new place to insert the Empiricla Validation of EMM

\subsection{Data Collection for Model Validation} \label{sec:data_validation}
To validate our Extended Mathematical Model against operational network behavior, we extracted three synchronized datasets from the VPS Servers running an I2P router at regular intervals: peer profile snapshots and active tunnel compositions. Profile data captured the router's local view of the network ($\omega_s$) along with measured speed and capacity values for visible peers. Tunnel data recorded the actual peer selections made during client and exploratory tunnel construction.

Data collection occurred over multiple observation windows, corresponding to I2P's tunnel rebuild cycle \cite{schimmer2009peer}. For each router $n_s$, we recorded: (1) the complete set of profiled peers with their tier classifications ($\gamma_s$, $g_{1s}$, $g_{2s}$), (2) peers selected for client tunnels ($\tau_s$), and (3) peers selected for exploratory tunnels ($\tau_{es}$). This dataset enabled direct comparison between predicted selection probabilities (Corollaries~\ref{corollary5} to \ref{corollary7}) and observed selection frequencies. It is used to validate specific model components: corollary~\ref{corollary5} validation utilized client tunnel selections to test Fast-tier preference with High Capacity fallback, corollary~\ref{corollary6} examined whether High Capacity peers achieved Fast tier membership according to threshold-based promotion logic, corollary~\ref{corollary7} tested capacity-based tier assignment across the complete router-visible peer population. The threshold-based classification mechanism formalized in Algorithm~\ref{alg:tier_classification} underpins all three validation exercises.

The validation data for Corollary~\ref{corollary5} comprised 
two synchronized components. Observation 1 totaled 76,687 peer profile measurements across ten temporal snapshots spanning minutes, capturing 8,260 distinct peers alongside their tier classifications and performance measures. In Observation 2, 63 client tunnel segment formations were captured during five of these snapshots, recording 45 distinct peers chosen as tunnel participants. The validation analysis involved the testing of 76 peers based on Corollary~\ref{corollary5}, where the observed and expected frequencies of client tunnel participation were compared against a predicted uniform distribution within the Fast tier. Goodness-of-fit Chi-square testing gave $p = 0.976$.

Validation of Corollary~\ref{corollary6} tracked 188 peers across the same ten time points, recording their tier classifications (Fast, High Capacity, or Standard) at each instance. This longitudinal design allowed testing whether High Capacity peers would become Fast tier members as per the threshold-based logic of Corollary~\ref{corollary6}. The result of the Chi-square comparison between observed and predicted conditional tier membership gave $p = 1.0$.

The validation of Corollary~\ref{corollary7} was conducted using a single-timestamp snapshot where 11,659 capacity measurements were obtained from 11,654 distinct peers visible in the router's Network Database. In the case of each peer, the logged dataset contained measured capacity values allowing a direct test of Corollary~\ref{corollary7}'s prediction: that peers exceeding the mean capacity and falling within the Maximum High Capacity Set limit are classified as High Capacity. The Kolmogorov-Smirnov test comparing the empirical and theoretical cumulative distribution functions of tier membership provided a test statistic of zero. All validation datasets, statistical test implementations, and analysis notebooks are available in the research repository~\cite{siddique_i2pemm}.

\subsection{Predicted Emergent Network Structure}
\label{sec:predicted_topology}

The EMM formally establishes peer selection at individual routers, but has network-wide consequences. Every router constructs tunnels based on its local perspective, which represents about 24.8\% of the global peer population (Corollary~\ref{corollary3}, Fig.~\ref{fig:network_visibility}). Thus, a single measurement cannot observe the entire topology that results from these collective choices. The EMM fills this gap: by composing the selection probabilities derived from Lemma~\ref{lemma1}, Lemma~\ref{lemma2}, and Theorem~\ref{theorem_structure_4}, it predicts the global structure when $N_{\omega_{I2P}} \approx 50{,}000$ routers~\cite{muntaka2025mapping} each execute the implementation-grounded logic defined in Algorithm~\ref{alg:tier_classification}.

This predicted topology is rendered on a Poincaré disk, as seen in Fig.~\ref{fig:emm_topology}, which is in accordance with the hyperbolic geometry induced by the XOR-metric keyspace of I2P~\cite{timpanaro2012bird}. Every node represents an I2P router; size indicates tier membership ($\gamma_s \subseteq g_{1s} \subseteq \omega_s$) according to Algorithm~\ref{alg:tier_classification}. Colour encodes tier membership - Fast (amber/gold), High Capacity (cyan), Standard (coral), and arc widths are proportional to the Standard-tier population distributed across the disk. The predicted tunnel connections are based on Theorem~\ref{theorem_structure_4}.

Three structural predictions are formed. To begin with, Fast-tier connections prevail across the interior web. Theorem~\ref{theorem_structure_4} predicts a degree of $2\left(\tfrac{H \cdot N_{\omega_s}}{N_{\gamma_s}} + 1 + \beta(H+1)\right) \approx 1053$ 
for Fast routers, as opposed to $2(H+1) = 8$ for Standard routers 
($H = 3$, $N_{\omega_s} = 12{,}394$, $N_{\gamma_s} \leq 75$, $\beta = 7.4$). 
Although this numerically small tier is capped at a strict 
limit of $\mathit{MFS} = 75$
by the \texttt{ProfileOrganizer.java} codebase~\cite{muntaka2026systemic} 
(occasional snapshots of 76 peers in $\gamma_s$, observed in \cite{siddique_i2pemm}, 
are transient pre-demotion states arising when node churn adds a peer to $\gamma_s$ 
within the 351-second profiling cycle before the demotion pass in 
Algorithm~\ref{alg:tier_classification}, lines 21--24, executes and restores 
the ceiling to 75), this tier forms the structural core of the predicted global topology. Second, degree distribution is asymmetric across the network. This asymmetry is not designed but emerges from the non-uniform Fast-tier distribution, reflecting heterogeneous tier composition across routers (Section~\ref{sec:critique}, refutation of Assumption~2, Fig.~\ref{fig:speed_variation}). Third, and most critically, the predicted global 
structure is inaccessible to any participating router. Each router's decisions are 
constrained to its local $\omega_s$ (Corollaries~\ref{corollary1}--\ref{corollary3}), 
yet Theorem~\ref{theorem_structure_4} shows the aggregate converges toward the degree-stratified topology depicted here. This analytical bridge, from locally partial to globally predictable, distinguishes the EMM from both empirical characterisations, 
which describe observed fragments, and documentation-based models, whose assumptions, as demonstrated in 
Section~\ref{sec:critique}, diverge from implementation reality.

\section{Discussion}
\label{sec:discussion}

The four assumptions of the previous model were not failures of reasoning; each was sound and internally consistent given the available protocol documentation. What the experiments in this paper reveal is a deeper methodological distinction: documentation and deployed implementation are not epistemically equivalent sources of model parameters in complex adaptive systems. Complete visibility ($\Omega = \Omega_s$) is reasonable with a centrally designed system; it is not in I2P since the propagation of the NetDB is designed to be decentralized. The assumption of uniform tier composition ($\mathcal{F}_{s_1} = \mathcal{F}_{s_2}$) would be acceptable if profiling were objective; it is untrue because profiling is computed locally using relativistic measures of path-dependent speeds. The High Capacity fallback in client tunnel construction and the $7.4\times$ overrepresentation in exploratory tunnels are purely a matter of implementation routing logic, absent from high-level specifications.

These are not mere gaps in implementation that could be eliminated by a more precise specification. They develop a conclusive argument concerning the points at which the valid analytical basis regarding network science should start. The degree asymmetry highlighted by Theorem~\ref{theorem_structure_4} is open to comparison with the work by Barab\'{a}si and Albert~\cite{barabasi1999emergence}, and the contrast between the two is instructive since the classical model's assumptions fail in exactly the ways that I2P's architecture makes inevitable. Classical preferential attachment provides an unlimited power-law distribution. The I2P topology is categorically different: the $\mathit{MFS} = 75$ hard algorithmic ceiling separates node degree from historical growth time-scales to yield a hub-based distribution with a narrow elite tier at $\approx 1053$ and the rest of the mass at $8$. It is still true that the bigger network science concept holds: structurally dominant hubs may arise due to simple local rules without global coordination~\cite{barabasi1999emergence}. Nonetheless, this generative process is algorithmic selection with a given limit, not preferential attachment. The resilience effects are essentially different. Scale-free networks break apart quickly when a few hub nodes are removed, whereas Algorithm~\ref{alg:tier_classification} continually replenishes the Fast tier from the High Capacity pool even when a node fails. The seventy-five-node ceiling is simultaneously the most recognizable structural weakness of the network and one of its main adaptive advantages.

This topology, which is dominated by hubs, has concrete security implications. A Fast-tier router is involved in a mathematically disproportionate fraction of all tunnel segments. A rival who is able to propel a node under their control to this level of elite performance through Sybil-based performance manipulation~\cite{egger2013practical} achieves structural leverage that is orders of magnitude higher than their numerical presence. This adversarial dynamic parallels known vulnerabilities found in similar complex networks such as in Tor, where contributing high-bandwidth nodes directly buys disproportionate path selection probabilities~\cite{oluwadare2026bandwidth}. However, our model demonstrates that I2P's vulnerability is categorically different: rather than scaling continuously with bandwidth, adversarial power in I2P concentrates violently once a node breaches the mathematical ceiling ($\mathit{MFS}=75$) of the Fast tier. This vulnerability is enhanced by the fact that the $7.4\times$ High Capacity overrepresentation in exploratory tunnels acts as an important routing bridge between tunnel layers, a structural role which lies completely outside the view of documentation-based analysis. In the absence of a formal structural model quantifying this specific leverage, the network structures that dark web command-and-control infrastructures use~\cite{jeong2016longitudinal, yasui2024spot} cannot be mapped. Thus, the adversarial resource level demanded to live up to principled cyber deterrence~\cite{rid2015attributing} is not formally specifiable.

The visibility constraint of approximately $24.8\%$ introduces a serious dimension of complex systems theory. Individual routers construct tunnels based on a very incomplete picture of a network whose global topology is formed entirely by the sum of all these incomplete pictures. Therefore, no participant can see the structure that they are all collectively creating. As Anderson's foundational work on emergence in physical systems establishes, large collections of interacting elements exhibit macroscopic properties that are not visible at the level of any individual component and must be characterized at the scale at which they arise \cite{anderson1972more}: the global topology is an analytic product of locality-constrained peer-selection rules in a structure no individual router observes or gives birth to intentionally. The predicted aggregate structure exhibits classical Watts-Strogatz small-world characteristics~\cite{watts1998collective}: high local clustering among Standard-tier peers and shorter global paths crossing Fast-tier hubs of degree beyond $10^3$. This concurs with an established outcome by Freenet that local routing policies create globally efficient structures which do not need central coordination~\cite{bibley}. Neither of the two properties is modeled into the protocol documentation; both arise dynamically as analytical results of tier-stratified peer selection.

\section{Conclusion}
\label{conclusion}

This work presents the first implementation-grounded Extended Mathematical Model of the emergent network topology of the Invisible Internet Project. The work shows that the epistemic gap between design specification and actual implementation is structurally consequential, not just a documentation artifact, by deriving operational parameters out of tested codebase instead of interpreting idealized documentation. 

To extend the analytical model of Bou Abdo and Hossain~\cite{abdo2023modeling}, four documentation-based assumptions were empirically recalibrated against 
operational testbed data. Production routers (nodes) monitor about $24.8\%$ of the active network, resulting in a disproportion to the assumption of complete global visibility. The measurements of speeds for the same peers differed by as much as $141.12$~KBps geographically, making the tier compositions highly relativistic. A preference for client tunnel formation of $97.3\%$ Fast-tier with a quantifiable High Capacity fallback remedied the supposition of an exclusive Fast-tier. The Standard-first strategy was used in exploratory tunnel construction, which involved a $7.4\times$ algorithmic High Capacity peer overrepresentation, proving the incorrect assumption of proportional mixing at the tier level. When these parameters were incorporated into the Extended Mathematical Model, they were statistically validated with a value of $p=0.976$ and $p=1.0$ in Chi-square testing, alongside a Kolmogorov-Smirnov statistic of zero.

Within these validated implementation parameters, Theorem~\ref{theorem_structure_4} derives a degree asymmetry of $131.6\times$ between Fast-tier structural hubs ($\approx 1053$) and Standard routing peers ($= 8$). This extreme level of topological imbalance is not motivated by classical preferential attachment, but is determined by an algorithmic selection with constraints below a hard $\mathit{MFS}=75$ population ceiling. It is a hub-based, bounded-elite architecture serving as the structural backbone of a $50{,}000$-router complex system which is structurally invisible to any one router within it. Analytically deriving this emergent architecture is the principal contribution of this work to network science. The structural conditions governing I2P's anonymity guarantees and adversarial leverage thresholds are now formally characterizable for the first time, providing the analytical foundation necessary for principled assessment of I2P-based command-and-control infrastructure.

\subsection{Future Directions}
\label{sec:future}

The most immediate open question concerns the operational stability of the overrepresentation factor ($\beta=7.4$). This parameter drives the exploratory tunnel component of Theorem~\ref{theorem_structure_4} and was measured under specific testbed conditions. Should $\beta$ shift under severe network congestion or adversarial flooding, the degree prediction for Fast-tier nodes would need to express $\beta$ as a dynamic function of network conditions rather than a fixed scalar.

Corollary~\ref{corollary3} currently treats the $24.8\%$ visibility constraint as a uniform distribution. In practice, NetDB queries follow strict XOR distance gradients~\cite{timpanaro2012bird}, concentrating each router's visible population within its immediate Kademlia neighborhood. Incorporating this non-Euclidean geometric bias into the selection probability matrix would improve the precision of degree distribution predictions, particularly in cyber warfare scenarios where geographic and XOR-distance topologies diverge substantially.

Extending the model to the floodfill routing layer is a natural next step. Floodfill routers govern which peers enter the visibility space ($\omega_s$) of standard routers in the first place. Formalizing their election criteria and NetDB coverage patterns would complete a unified bottom-up account of I2P topology formation, from the database layer through to tunnel construction.

Two security applications of the EMM remain open. First, Theorem~\ref{theorem_structure_4} gives the exact probability that a Fast-tier node participates in any given tunnel construction event. From this, the fraction of traffic passively observable by an adversary controlling $k$ of the $\mathit{MFS}=75$ Fast tier nodes can be analytically derived, yielding a structural and quantitative definition of I2P's anonymity guarantee as a direct function of adversarial penetration depth. Second, because Fast-tier membership demands sustained above-median performance against a hard population ceiling, the infrastructure investment required to reach a target observation rate is mathematically bounded and quantifiable. The same model extends naturally to defensive applications, allowing law enforcement to quantify the minimum resource investment needed to achieve a target disruption rate against malicious infrastructure operating within I2P. This economic dimension is precisely what the cyber deterrence literature~\cite{rid2015attributing} demands for credible threat assessment and principled resource allocation. The methodology is directly applicable to Freenet, GNUnet, and other anonymity systems.

\ifCLASSOPTIONcompsoc



%\bibliographystyle{IEEEtran}

% ============================================================
% REFERENCES
% ============================================================
\bibliographystyle{IEEEtran}
\bibliography{ref}

@inproceedings{abdo2023modeling,
  title={Modeling the Invisible Internet},
  author={Bou Abdo, Jacques and Hossain, Liaquat},
  booktitle={International Conference on Complex Networks and Their Applications},
  pages={359--370},
  year={2023},
  organization={Springer}
}

@article{de2019invisible,
  title={Invisible Internet Project (I2P)},
  author={De Boer, Tim and Breider, Vincent},
  journal={System and Network Engineering},
  pages={1--16},
  year={2019}
}

@inproceedings{herrmann2011privacy,
  title={Privacy-Implications of Performance-Based Peer Selection by Onion-Routers: A Real-World Case Study Using {I2P}},
  author={Herrmann, Michael and Grothoff, Christian},
  booktitle={International Symposium on Privacy Enhancing Technologies Symposium},
  pages={155--174},
  year={2011},
  organization={Springer}
}

@inproceedings{hoang2018empirical,
  title={An Empirical Study of the {I2P} Anonymity Network and Its Censorship Resistance},
  author={Hoang, Nguyen Phong and Kintis, Panagiotis and Antonakakis, Manos and Polychronakis, Michalis},
  booktitle={Proceedings of the Internet Measurement Conference 2018},
  pages={379--392},
  year={2018}
}

@inproceedings{liu2019i2p,
  title={{I2P} Anonymous Communication Network Measurement and Analysis},
  author={Liu, Likun and Zhang, Hongli and Shi, Jiantao and Yu, Xiangzhan and Xu, Haixiao},
  booktitle={Smart Computing and Communication: 4th International Conference, SmartCom 2019},
  pages={105--115},
  year={2019},
  organization={Springer}
}

@inproceedings{schimmer2009peer,
  title={Peer Profiling and Selection in the {I2P} Anonymous Network},
  author={Schimmer, Lars},
  booktitle={PetCon 2009.1},
  pages={59--70},
  year={2009},
  organization={Technische Universit{\"a}t Dresden}
}

@inproceedings{yin2019i2p,
  title={{I2P} Anonymous Traffic Detection and Identification},
  author={Yin, Hongshan and He, Yongzhong},
  booktitle={2019 5th International Conference on Advanced Computing \& Communication Systems (ICACCS)},
  pages={157--162},
  year={2019},
  organization={IEEE}
}

@inproceedings{egger2013practical,
  title={Practical Attacks Against the {I2P} Network},
  author={Egger, Christoph and Schlumberger, Johannes and Kruegel, Christopher and Vigna, Giovanni},
  booktitle={Research in Attacks, Intrusions, and Defenses: 16th International Symposium, RAID 2013},
  pages={432--451},
  year={2013},
  organization={Springer}
}

@inproceedings{borisov2007denial,
  title={Denial of Service or Denial of Security?},
  author={Borisov, Nikita and Danezis, George and Mittal, Prateek and Tabriz, Parisa},
  booktitle={Proceedings of the 14th ACM Conference on Computer and Communications Security},
  pages={92--102},
  year={2007}
}

@misc{Maor2014,
  author={Maor, Etay},
  title={Out of the Shadows: {i2Ninja} Malware Exposed},
  howpublished={Security Intelligence},
  url={https://securityintelligence.com/shadows-i2ninja-malware-exposed},
  year={2014}
}

@inproceedings{jeong2016longitudinal,
  title={A Longitudinal Analysis of {.i2p} Leakage in the Public {DNS} Infrastructure},
  author={Jeong, Seong Hoon and Kang, Ah Reum and Kim, Joongheon and Kim, Huy Kang and Mohaisen, Aziz},
  booktitle={Proceedings of the 2016 ACM SIGCOMM Conference},
  pages={557--558},
  year={2016}
}

@article{yasui2024spot,
  title={{SPOT}: In-Depth Analysis of {IoT} Ransomware Attacks Using Bare Metal {NAS} Devices},
  author={Yasui, Hiroki and Inoue, Takahiro and Sasaki, Takayuki and Tanabe, Rui and Yoshioka, Katsunari and Matsumoto, Tsutomu},
  journal={Journal of Information Processing},
  volume={32},
  pages={23--34},
  year={2024},
  publisher={Information Processing Society of Japan}
}

@article{chao2024systematic,
  title={A Systematic Survey on Security in Anonymity Networks: Vulnerabilities, Attacks, Defenses, and Formalization},
  author={Chao, Daichong and Xu, Dawei and Gao, Feng and Zhang, Chuan and Zhang, Weiting and Zhu, Liehuang},
  journal={IEEE Communications Surveys \& Tutorials},
  volume={26},
  number={3},
  pages={1775--1829},
  year={2024},
  publisher={IEEE}
}

@article{muntaka2025mapping,
  author    = {Muntaka, S. A. and Bou Abdo, J. and Akanbi, K. and Oluwadare, S. and Hussein, F. and Kornyo, O. and Asante, M.},
  title     = {Mapping The Invisible Internet: Framework and Dataset},
  journal   = {Data in Brief},
  volume    = {63},
  pages     = {112175},
  year      = {2025},
  month     = oct,
  doi       = {10.1016/j.dib.2025.112175}
}

@article{muntaka2025resilience,
  title={Resilience of the Invisible Internet Project: A Computational Analysis},
  author={Muntaka, Siddique Abubakr and Bou Abdo, Jacques},
  journal={Internet Technology Letters},
  volume={8},
  number={5},
  pages={e70119},
  year={2025},
  publisher={Wiley Online Library}
}

@article{barabasi1999emergence,
  title={Emergence of Scaling in Random Networks},
  author={Barab{\'a}si, Albert-L{\'a}szl{\'o} and Albert, R{\'e}ka},
  journal={Science},
  volume={286},
  number={5439},
  pages={509--512},
  year={1999},
  publisher={American Association for the Advancement of Science}
}

@inproceedings{maymounkov2002kademlia,
  title={Kademlia: A Peer-to-Peer Information System Based on the {XOR} Metric},
  author={Maymounkov, Petar and Mazieres, David},
  booktitle={International Workshop on Peer-to-Peer Systems},
  pages={53--65},
  year={2002},
  organization={Springer}
}

@article{dingledine2004tor,
  title={Tor: The Second-Generation Onion Router},
  author={Dingledine, Roger and Mathewson, Nick and Syverson, Paul},
  year={2004}
}

@inproceedings{murdoch2005low,
  title={Low-Cost Traffic Analysis of {Tor}},
  author={Murdoch, Steven J and Danezis, George},
  booktitle={2005 IEEE Symposium on Security and Privacy (S\&P'05)},
  pages={183--195},
  year={2005},
  organization={IEEE}
}

@article{muntaka2025optimizing,
  title={Optimizing Anonymity and Efficiency: A Critical Review of Path Selection Strategies in {Tor}},
  author={Muntaka, Siddique Abubakr and Bou Abdo, Jacques},
  journal={arXiv preprint arXiv:2508.17651},
  year={2025}
}

@inproceedings{timpanaro2012bird,
  title={A Bird's Eye View on the {I2P} Anonymous File-Sharing Environment},
  author={Timpanaro, Juan Pablo and Chrisment, Isabelle and Festor, Olivier},
  booktitle={International Conference on Network and System Security},
  pages={135--148},
  year={2012},
  organization={Springer}
}

@phdthesis{timpanaro2011monitoring,
  title={Monitoring the {I2P} Network},
  author={Timpanaro, Juan Pablo and Isabelle, Chrisment and Olivier, Festor},
  year={2011},
  school={Inria}
}

@misc{siddique_i2pemm,
  author = {Muntaka, Siddique Abubakr},
  title = {{I2PEMM}: {I2P} Extended Mathematical Modelling Framework for Network Topology Analysis},
  year = {2026},
  url = {https://github.com/abksiddique/I2PEMM},
  note = {GitHub Repository}
}

@online{bibley,
  author       = {{Freenet Project Inc.}},
  title        = {P2P Network},
  year         = {2024},
  url          = {https://freenet.org/resources/manual/architecture/p2p-network/},
  note         = {[Online]}
}

@misc{i2p_peermanager_source,
  author={{Invisible Internet Project (I2P)}},
  title={PeerManager.java: Peer Profile Reorganization Constants},
  year={2024},
  url={https://github.com/i2p/i2p.i2p/blob/18b2b1faf9933fec89ec146c27bd6bdb9fe3ae05/router/java/src/net/i2p/router/peermanager/PeerManager.java#L48-L56},
  note={GitHub Repository. Lines 48--56 define REORGANIZE\_TIME (45s), REORGANIZE\_TIME\_MEDIUM (123s), and REORGANIZE\_TIME\_LONG (351s) constants for profile update scheduling based on router uptime}
}

@article{muntaka2026systemic,
  author  = {Muntaka, Siddique Abubakr and Bou Abdo, Jacques},
  title   = {Systemic Flaws in the Invisible Internet Project: Analysis of Exploitable Design Choices},
  journal = {Authorea Preprints},
  year    = {2026},
  month   = feb,
  doi     = {10.22541/au.177031274.48275457/v1},
  url     = {https://doi.org/10.22541/au.177031274.48275457/v1}
}

@article{chaum1981untraceable,
  title={Untraceable electronic mail, return addresses, and digital pseudonyms},
  author={Chaum, David L},
  journal={Communications of the ACM},
  volume={24},
  number={2},
  pages={84--90},
  year={1981},
  publisher={ACM New York, NY, USA}
}

@article{rid2015attributing,
  title={Attributing cyber attacks},
  author={Rid, Thomas and Buchanan, Ben},
  journal={Journal of strategic studies},
  volume={38},
  number={1-2},
  pages={4--37},
  year={2015},
  publisher={Taylor \& Francis}
}

@article{anderson1972more,
  title={More is different: broken symmetry and the nature of the hierarchical structure of science.},
  author={Anderson, Philip W},
  journal={Science},
  volume={177},
  number={4047},
  pages={393--396},
  year={1972},
  publisher={American Association for the Advancement of Science}
}

@article{watts1998collective,
  title={Collective dynamics of ‘small-world’networks},
  author={Watts, Duncan J and Strogatz, Steven H},
  journal={nature},
  volume={393},
  number={6684},
  pages={440--442},
  year={1998},
  publisher={Nature Publishing Group}
}

@article{euler1741solutio,
  title={Solutio problematis ad geometriam situs pertinentis},
  author={Euler, Leonhard},
  journal={Commentarii academiae scientiarum Petropolitanae},
  pages={128--140},
  year={1741}
}

@article{erdos1959publicationes,
  title={Publicationes mathematicae debrecen},
  author={Erd{\"o}s, P and R{\'e}nyi, A and Bollob{\'a}s, B},
  journal={Random Graphs I},
  volume={6},
  pages={290--297},
  year={1959}
}

@inproceedings{georgoulias2023market,
  title={In the market for a Botnet? An in-depth analysis of botnet-related listings on Darkweb marketplaces},
  author={Georgoulias, Dimitrios and Pedersen, Jens Myrup and Hutchings, Alice and Falch, Morten and Vasilomanolakis, Emmanouil},
  booktitle={2023 APWG Symposium on Electronic Crime Research (eCrime)},
  pages={1--14},
  year={2023},
  organization={IEEE}
}

@misc{cyberpress_ratatouille_i2p_2026,
  author       = {{CyberPress}},
  title        = {Ratatouille Malware Exploits {I2P} Network to Bypass {UAC} Control},
  year         = {2026},
  url          = {https://cyberpress.org/ratatouille-malware-exploits-i2p-network-bypass-uac-control/},
  note         = {Online news article. Accessed: February 2026}
}

@article{oluwadare2026bandwidth,
  title={When Bandwidth Buys Power: Quantifying Anonymity in Tor},
  author={Oluwadare, Sunkanmi and Bou Abdo, Jacques},
  journal={Authorea Preprints},
  year={2026},
  publisher={Authorea}
}

@article{wang2025time,
  title={Time will Tell: Large-scale De-anonymization of Hidden I2P Services via Live Behavior Alignment (Extended Version)},
  author={Wang, Hongze and Ling, Zhen and Xu, Xiangyu and Pan, Yumingzhi and Liu, Guangchi and Luo, Junzhou and Fu, Xinwen},
  journal={arXiv preprint arXiv:2512.15510},
  year={2025}
}

@article{muntaka2026fifty,
  title={Fifty Shades of Darknet},
  author={Muntaka, Siddique Abubakr and Abdo, Jacques Bou},
  journal={arXiv preprint arXiv:2605.19437},
  year={2026}
}

% ============================================================
% AUTHOR BIOGRAPHIES
% ============================================================

\begin{IEEEbiography}[{\includegraphics[width=1in,height=1.25in,clip,keepaspectratio]{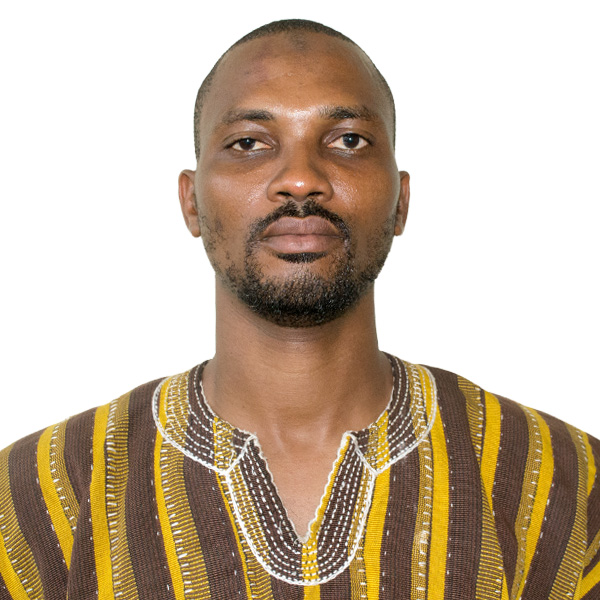}}]{Siddique Abubakr Muntaka}
received the M.Sc. degree in information technology from Kwame Nkrumah
University of Science and Technology, Kumasi, Ghana, in 2019. He is currently
pursuing the Ph.D. degree in information technology with the School of
Information Technology, University of Cincinnati, Cincinnati, OH, USA, where
he also serves as a Graduate Research Assistant, Teaching Assistant, and
Adjunct Faculty. He has more than a decade of professional experience in
systems and network engineering, having served as a Lecturer and I.T. Manager
at Kessben University College, Kumasi, and as a part-time Lecturer at
Wisconsin International University College, Ghana. His research interests
include anonymity networks (I2P and Tor), network science, complex systems,
cloud infrastructure, and cybersecurity. His recent work on mapping and
modeling the Invisible Internet Project has appeared in Elsevier, \textit{Internet
Technology Letters} and IEEE conference proceedings.
\end{IEEEbiography}

\begin{IEEEbiography}[{\includegraphics[width=1in,height=1.25in,clip,keepaspectratio]{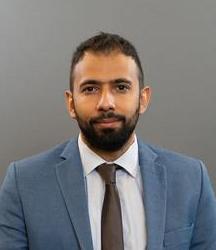}}]{Jacques Bou Abdo}
received the Ph.D. degree in computer science, with a focus on cybersecurity,
from Sorbonne University, Paris, France, and the Ph.D. degree in management
sciences, with a focus on network economics, from Universit\'e Paris-Saclay,
Paris, France. He is currently an Assistant Professor with the School of
Information Technology, University of Cincinnati, Cincinnati, OH, USA. He is
an interdisciplinary researcher whose research interests include complex
systems, cybersecurity, cyber warfare, anonymity networks, computational
economics, and network economics.
\end{IEEEbiography}

\begin{IEEEbiography}[{\includegraphics[width=1in,height=1.25in,clip,keepaspectratio]{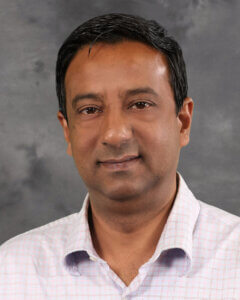}}]{Liaquat Hossain}
received the M.Sc. degree in computer and engineering management from
Assumption University, Bangkok, Thailand, in 1995, and the Ph.D. degree in
information technology and computer science from the University of
Wollongong, Wollongong, NSW, Australia, in 1997. He conducted postdoctoral
research with the Internet Telephony Interoperability Consortium,
Massachusetts Institute of Technology, Cambridge, MA, USA. He is currently
the Dean and a Professor with the College of Science, Long Island University,
Brooklyn, NY, USA. He has previously held senior academic leadership roles
including Associate Dean at The University of Sydney, Australia, and Head of
Division and Associate Dean of Research at The University of Hong Kong, Hong
Kong. His research interests include network science, complex systems,
coordination in large-scale networks, resilient systems, and public health
preparedness.
\end{IEEEbiography}

\vfill
\end{document}